\documentclass[twocolumn,aps,prl]{revtex4-2}

\usepackage{subfig}
\usepackage{graphicx}
\usepackage{amsmath, xparse}
\usepackage{xcolor}
\usepackage{bbm}
\usepackage{amsfonts}
\usepackage{mathtools}

\begin{document}
	\title{Exactly solving the Kitaev chain and generating Majorana-zero-modes out of noisy qubits}
	
	\author{Marko J. Ran\v{c}i\'{c}}
	
	\email{marko.rancic@totalenergies.com}
	
	\address{Address TotalEnergies, Tour Coupole La Défense, 2 Pl. Jean Millier, 92078 Paris}
	
	\begin{abstract}
		\section{Abstract}
		Majorana-zero-modes (MZMs) were predicted to exist as edge states of a physical system called the Kitaev chain. MZMs should host particles that are their own antiparticles and could be used as a basis for a qubit which is robust-to-noise. However, all attempts to prove their existence gave inconclusive results. Here, the Kitaev chain is exactly solved with a quantum computing methodology and properties of MZMs are probed by generating eigenstates of the Kitev Hamiltonian on 3 noisy qubits of a publicly available quantum computer. After an ontological elaboration I show that two eigenstates of the Kitaev Hamiltonian exhibit eight signatures attributed to MZMs. The results presented here are a most comprehensive set of validations of MZMs ever conducted in an actual physical system. Furthermore, the findings of this manuscript are easily reproducible for any user of publicly available quantum computers, solving another important problem of research with MZMs - the result reproducibility crisis. 
	\end{abstract}

	\maketitle
	
	\section{Introduction}
	A Majorana fermion is its own antiparticle. This concept originally proposed in the context of particle physics more than 80 years ago experienced a rebirth with the work of Alexei Kitaev in early 2000s \cite{kitaev2001unpaired}. Kitaev proposed a toy model composed out of a chain of spinless fermions which are coupled by tunnelling $t$ in the presence of $p$-wave superconducting pairing $\Delta$ and tunable chemical potential $\mu$. A Kitaev chain has an exceptional feature that at low chemical potential and when tunnelling is comparable to the superconducting pairing exact zero energy solutions localised at the edges exist. Such states are immune to any small changes in local parameters and could potentially serve as a basis for a topological qubit \cite{kitaev2001unpaired,beenakker2013search,sarma2015majorana,PhysRevB.92.165118,PhysRevB.98.155414,PhysRevB.101.075404}
	
	Thus far, two common ways of getting a theoretical insight into the physics of Kitaev Hamiltonians in a broad range of parameters existed: numerical diagonalization of the many-body Hamiltonian and single-particle picture theories, such as diagonalizing the non-interacting Bogoliubov-de Gennes Hamiltonian. Numerical diagonalization of the many-body Hamiltonian is practically unfeasible for longer chains as the Hilbert space has $2^n$ states, where $n$ is the number of sites in the chain. On the other hand, the Bogoliubov-de Gennes Hamiltonian operates on a Hilbert space of $2n$ states and has two zero energy solutions in the topological regime at low chemical potential. However, this Hamiltonian is a single-particle one and solves the Kitaev chain in a mean-field flavour. The reader should also be referred to studies with Matrix-product states \cite{miao2018majorana} and Quantum Monte Carlo \cite{PhysRevB.104.035116}.
	
	Many theoretical proposals and experimental validations showing Majorana-like features followed after Kitaev's work \cite{PhysRevLett.106.220402,wiedenmann20164pi,PhysRevLett.100.096407,deng2012anomalous,PhysRevB.79.161408,rokhinson2012fractional,sau2012realizing,he2017chiral,banerjee2016proximate,wang2018evidence,manna2020signature,PhysRevLett.122.126402,hoffman2021majorana,PhysRevB.104.214501}. Some of the works received broad attention from the scientific community such as the spin-orbit nanowire in the presence of an external magnetic field and proximitized superconductivity \cite{mourik2012signatures} and deposited iron atoms on top of a superconductor \cite{nadj2014observation}. Nevertheless, all of these experimental validations were followed by theory or experiments of topologically trivial phenomena mimicking those of Majorana zero modes \cite{PhysRevLett.109.237009,ruby2017exploring,PhysRevB.96.075161,PhysRevB.97.214502,PhysRevLett.115.127003,lee2014spin,PhysRevLett.123.107703,PhysRevB.100.125407,vuik2019reproducing,prada2020andreev,kayyalha2020absence,kim2021anisotropic,chen2018discrete,jack2021detecting,chatzopoulos2021spatially}. To date, the verification of of MZMs remains one of the most debated on topics in physics.
	
	Noisy quantum computers of today represent versatile platforms for probing quantum properties of chemical systems \cite{peruzzo2014variational}, QED systems \cite{stetina2022simulating} and even exotic, previously not-realized, condensed matter states such as time crystals \cite{PRXQuantum.2.030346}, just to name a few. This manuscript aims to connect the world of MZMs with quantum computers. Here, I will present a method of solving the Kitaev chain Hamiltonian exactly with a quantum computing methodology and use a quantum computer to prepare exact eigenstates of such a Hamiltonian in what is commonly refereed to as a topologically non-trivial regime. The Kitaev chain in this study is composed out of noisy qubits. One key question which the reader of this study might have is: are the Majorana-zero-modes (MZMs) ''artificiality" created on a quantum computer within this study actual MZMs or a mere representation of MZMs on a quantum computer? Given that quantum computers prepare actual quantum-mechanical wavefunctions this question might be paraphrased as: if two wavefunctions are exactly the same do they describe the same physical reality? The question of the meaning of the wavefunction is almost as old of quantum mechanics itself as it was initially posed by Max Born Ref. \cite{born1926quantenmechanik}. Until recently the meaning of the wavefunction remained a somewhat disputed question with works of Ref. \cite{pusey2012reality} and Ref. \cite{PhysRevLett.108.150402} providing the most complete answer to-date. The author of this manuscript is closest to the view of Colbeck and Renner that a 1-to-1 mapping between the wavefunction and reality exist with a prior assumption that measurement settings could be freely chosen Ref. \cite{PhysRevLett.108.150402}. To put it simply: the MZMs generated on a quantum computer would be as real as they would be in any realisation in a condensed matter setting under the assumption of a non-super-deterministic universe.
	
	Recent studies with quantum computers have focused on a single trait of MZMs (predominantly braiding) at $\mu=0$ and $t=\Delta$ with approximate methods to prepare the ground state such as the imaginary time evolution. Historically the idea to probe braiding of MZMs with quantum computers was originally proposed in the context of superconducting qubits in Ref. \cite{you2014encoding} and was realised with a photonic system in Ref. \cite{xu2016simulating} and a superconducting system \cite{PhysRevResearch.3.033171}. Novel works with qubits also focus on teleportation of MZMs \cite{PhysRevLett.126.090502} and entanglement entropy \cite{xiao2021determining}.
	
	In contrast to the above-mentioned works here I will present a general methodology to exactly obtain the eigenstates of the Kitaev Hamiltonian on a quantum computer in a broad range of parameters $\mu$, $t$ and $\Delta$. Instead on focusing solely on a handful of features indicative of MZMs, my robust framework allows me to simultaneously test a record number of prediction about MZMs in an actual physical system. Two eigenstates of a 3-site Kitaev chain will show eight distinct features of MZMs: i) a robust to noise degeneracy with their ground states at low chemical potential, an important feature of MZMs as discussed in Ref. \cite{klassen2015topological}; ii) Upon visually comparing the measured spectrum with a classically obtained Bogoliubov-de-Gennes (BdG) single-particle spectrum of MZMs a striking similarity is observed, like predicted in Kitaev's original proposal \cite{kitaev2001unpaired}. This similarity is quantified by calculating a mean absolute error of the measured data with respect to the BdG predictions;
	iii) The states under study have a well defined parity as discussed in Ref. \cite{sarma2015majorana}, with iv) parity switches at specific values of the chemical potential in striking accordance with single-particle theories of MZMs - a feature originally predicted in Ref. \cite{hegde2015quench}; v) A non-conserved particle number of MZMs states as elaborated in Ref. \cite{lin2018towards}; vi) A Majorana-edge correlation function which decays with the chemical potential. Although this feature was quantitatively predicted in the thermodynamic limit here due to a finite chain size I display only qualitative matching \cite{miao2018majorana}; vii) A display that MZMs favour exclusively Majorana-edge pairing at low chemical potential and viii) nearest-neighbour pairing at large values of the chemical potential another feature predicted in Kitaev's  original proposal \cite{kitaev2001unpaired}.
	
	The goal of this study is to go beyond showing a handful of indications of MZMs and present a large number (eight) of corroborating evidence in a single reproducible experiment. Furthermore, the findings of this manuscript are easily reproducible for any user of publicly available quantum computers, solving another important issue with MZMs - the reproducibility crisis \cite{frolov2021quantum}. This study focuses on a $3$-site Kitaev chain for doing experiments on actual quantum computers and  $4$-site chains for simulating the results of noiseless quantum computers with Qiskit. The chain length in this study was limited by noise levels of quantum computers. Experiments with longer chains of up to $7$ qubits were a focus of a followup study \cite{sung2022preparing} performed together with researchers from IBM where error mitigation had to be applied to get results for longer chains because of the noise levels in contemporary quantum computers.
	
	\begin{figure*}
		\includegraphics[width=0.82\textwidth]{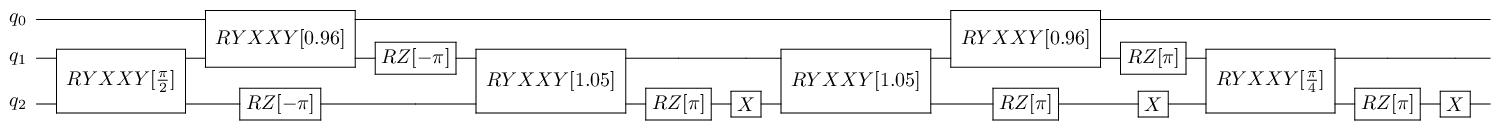}
		\caption{The circuit which generates a ground state of a 3 site Kitaev Hamiltonian at $t=-1$, $\Delta=1$ and $\mu=10^{-8}$. A general procedure for obtaining quantum computing circuits which represent the ground state of arbitrary quadratic Hamiltonians is given in Supplementary material Section S1.\label{fig:circ}}
	\end{figure*}
	
	\begin{figure*}
		\begin{minipage}{0.33\textwidth}
			\centering  
			\subfloat{(a)}{		
				
				\includegraphics[height=3.3cm]{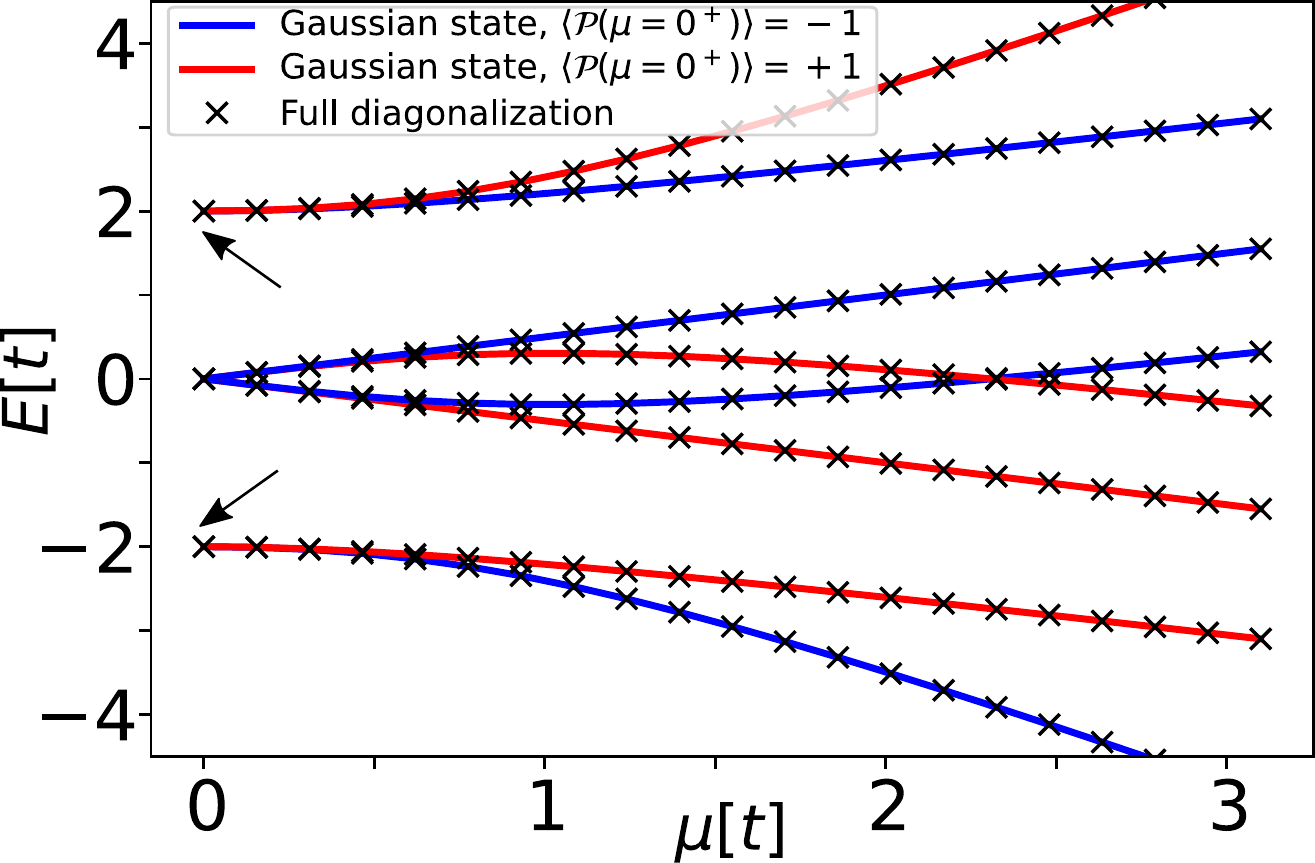}}
		\end{minipage}\begin{minipage}{0.33\textwidth}
			\centering
			\subfloat{(b)}{
				
				\includegraphics[width=1.0\textwidth]{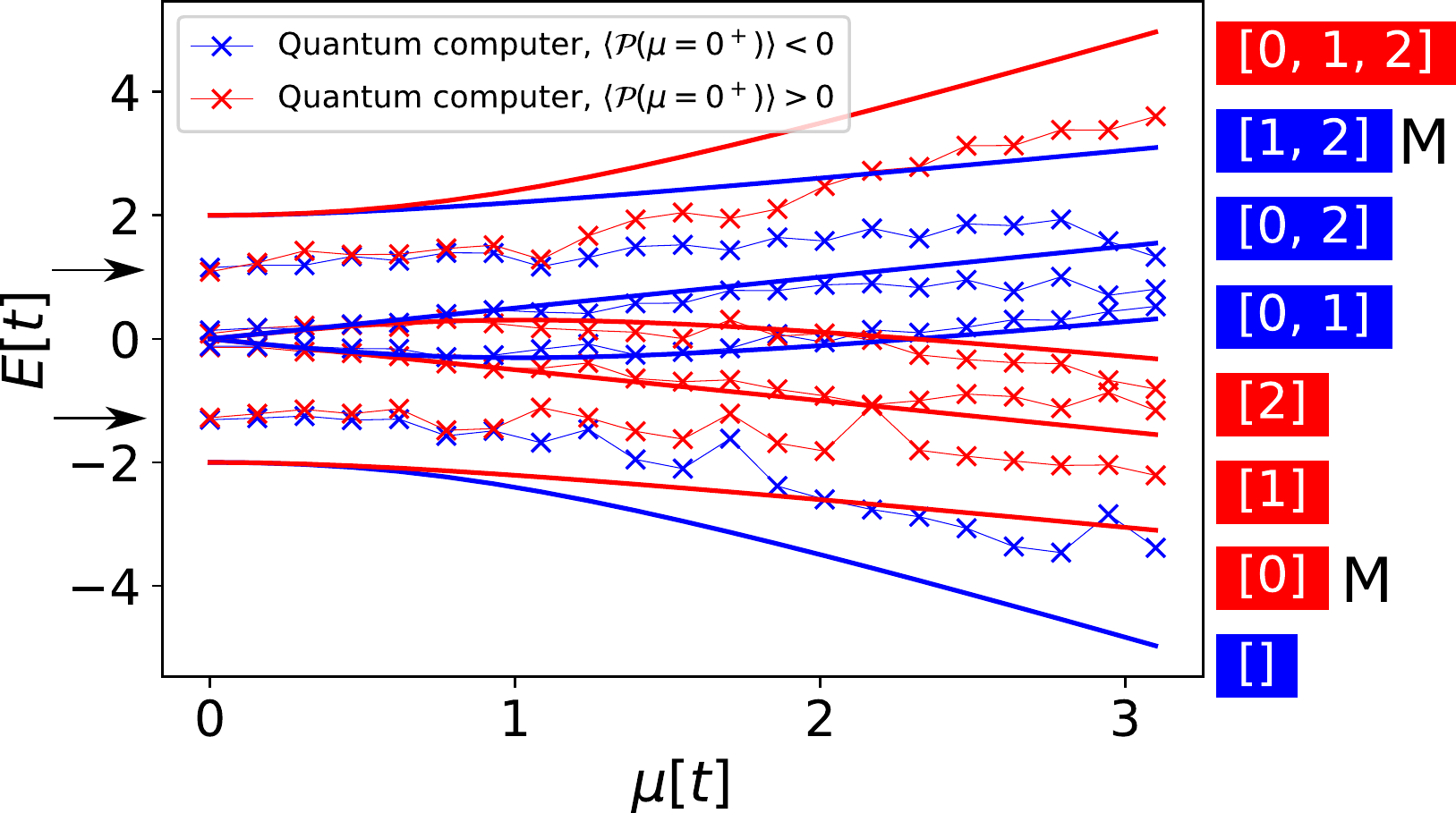}}
		\end{minipage}\begin{minipage}{0.33\textwidth}
			\centering
			\subfloat{(c)}{
				
				\includegraphics[height=3.3cm]{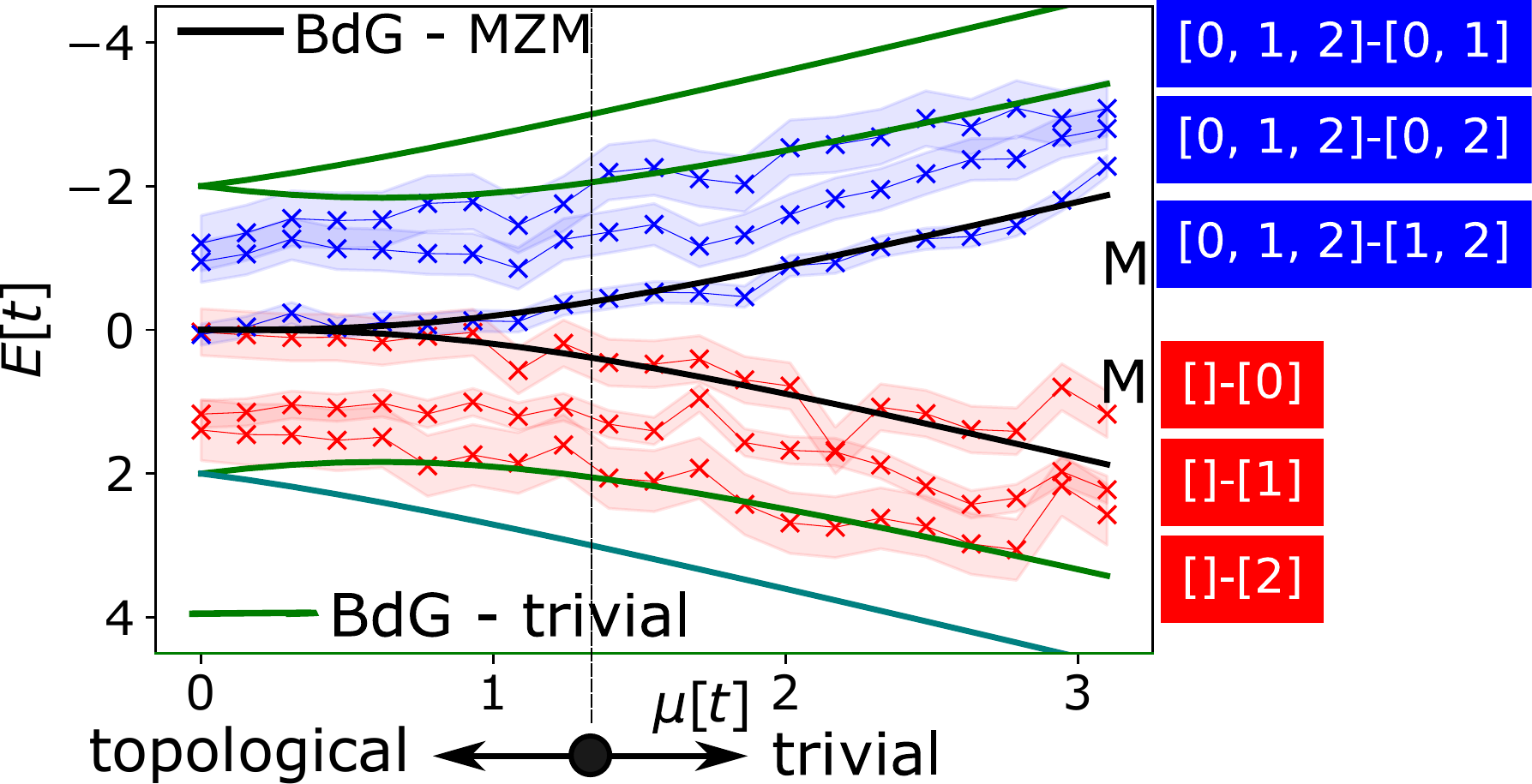}}
			
		\end{minipage}
		\begin{minipage}{0.33\textwidth}
			\centering
			\subfloat{(d)}{
				\includegraphics[width=1.0\textwidth]{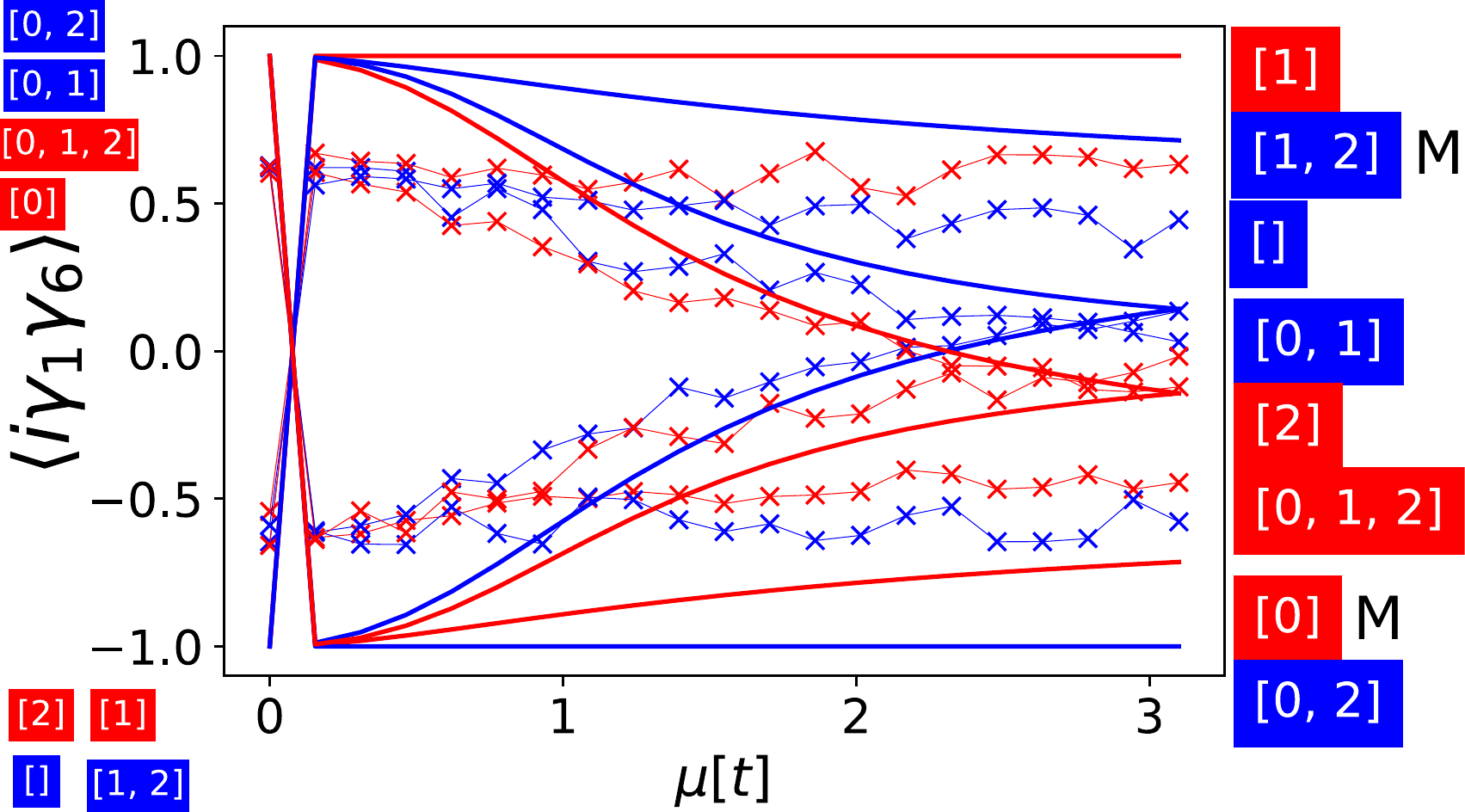}}
		\end{minipage}\begin{minipage}{0.33\textwidth}
			\centering
			
			\subfloat{(e)}{
				
				\includegraphics[height=3.4cm]{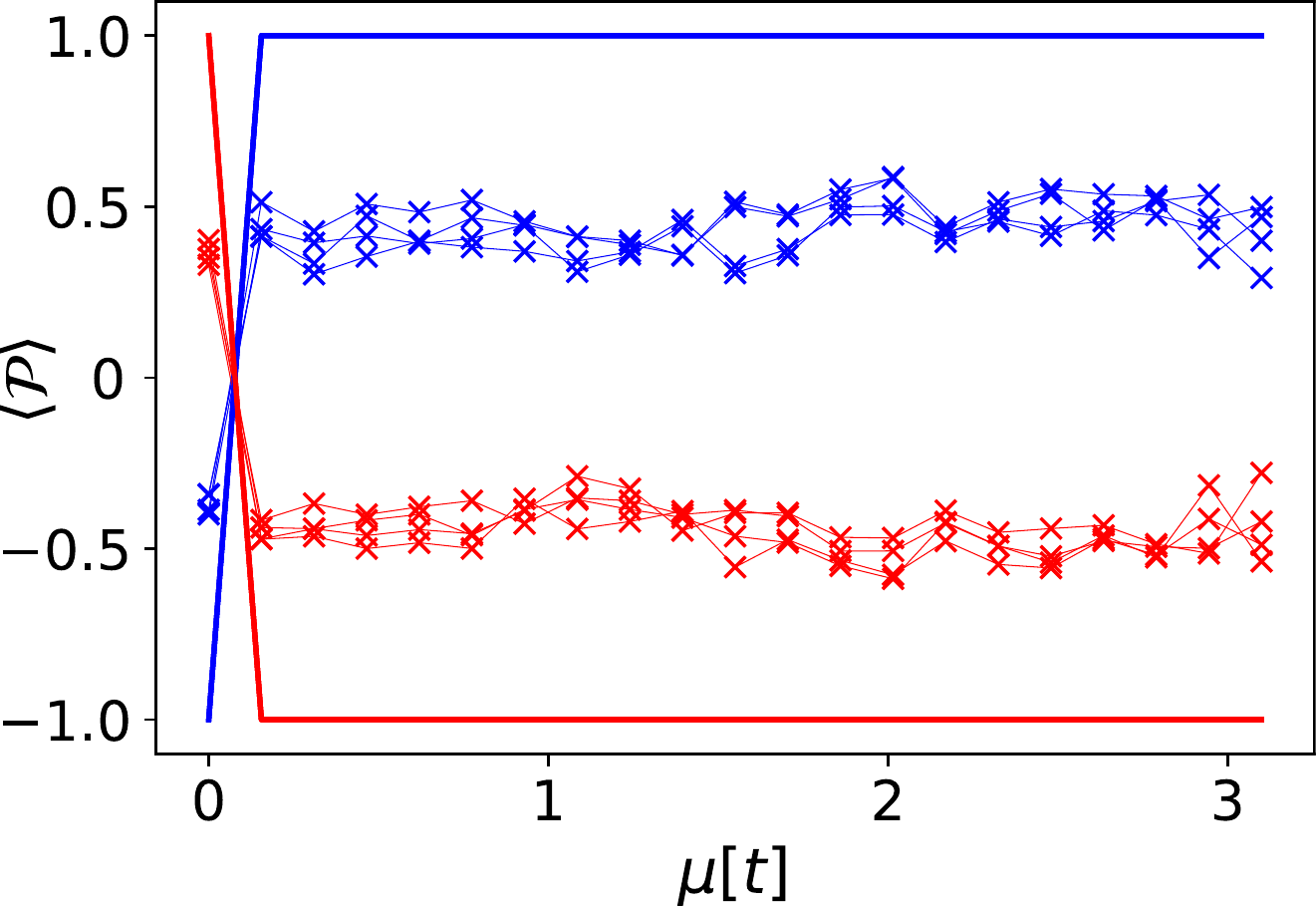}}
		\end{minipage}\begin{minipage}{0.33\textwidth}
			\centering
			\subfloat{(f)}{
				\includegraphics[width=0.95\textwidth]{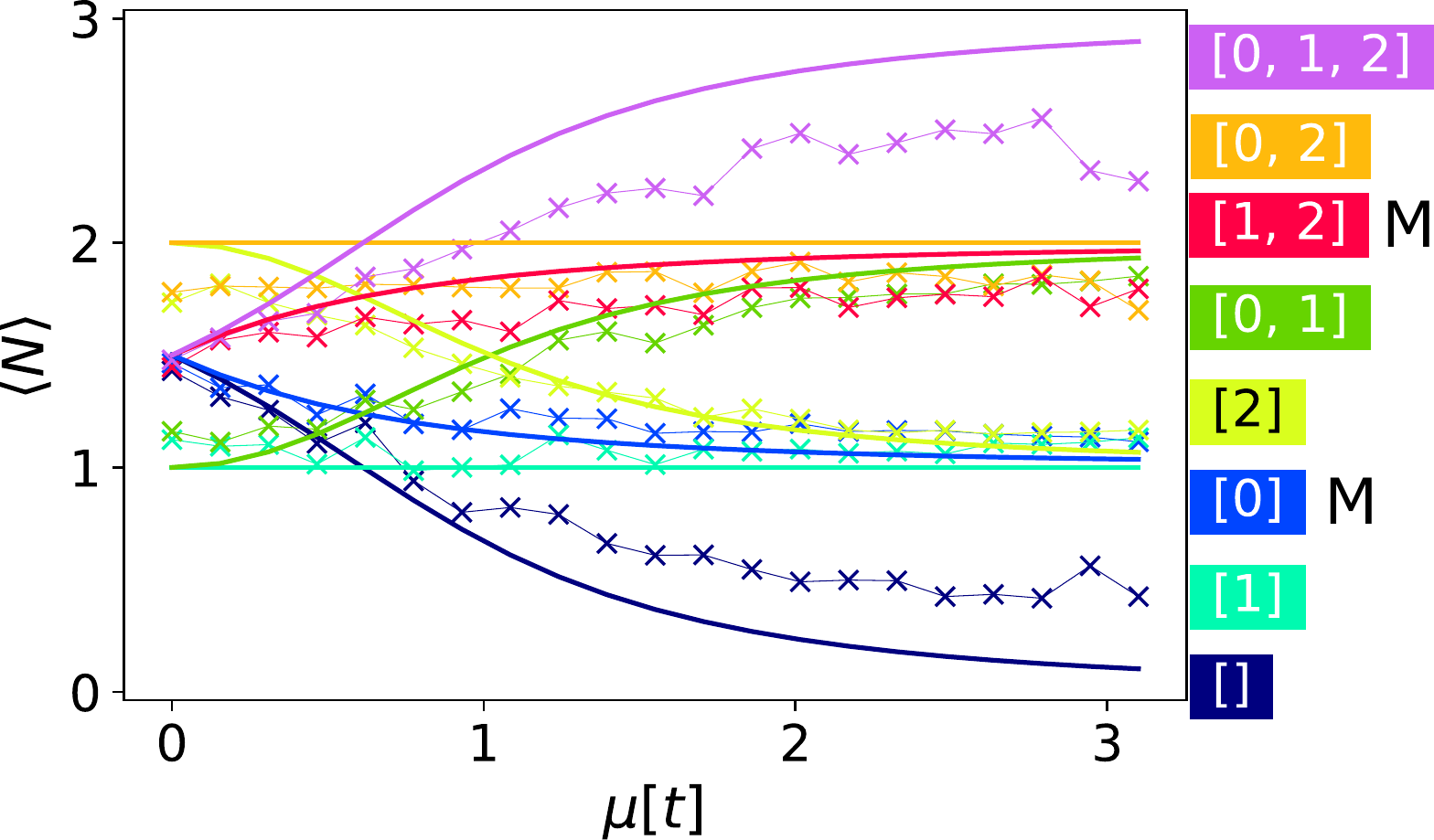}}
		\end{minipage}
		\caption{Quantities of the Kitaev Hamiltonian at $t=-1$ and $\Delta=1$ as a function of the chemical potential $\mu$ in units of absolute value of tunnelling $[t]$. (a-e) Black "x" symbols represent a result of a numerical diagonalization on a classical computer. Red "x" symbols (full lines) values obtained on a quantum computer (ideal simulator of quantum computers) with $\langle P(\mu=0^+)\rangle=1$. Blue "x" symbols (full lines) values obtained on a quantum computer (ideal simulator of quantum computers) with $\langle P(\mu=0^+)\rangle=-1$. The black arrows in (a-b) denote the position of the possible topological degeneracy. The "M" in (b), (c), (d) and (f) denotes the alleged MZM state. (a-b) The full spectrum of the Kitaev Hamiltonian obtained with Gaussian states compared to a numerical diagonalization (a) and IBMQ Santiago (b).  (c) The single-particle picture BdG spectrum (full line) compared to the BdG spectrum on IBMQ Santiago. Shaded blue and red regions represent a root-mean-squared deviation of a given state with respect to a theoretical prediction of the BdG spectrum on QPUs in presence of pure-dephasing (see Supplementary section S3 for more details about the noise model).  (d) Majorana edge-correlation function, an ideal simulation compared to the actual 3-site Kitaev chain composed out of qubits. (e) Parity - an ideal simulation compared to the actual 3-site Kitaev chain composed out of qubits. (f) Particle number on an ideal simulator (full lines) and actual 3-site Kitaev chain composed out of qubits (x-shaped coloured symbols). \label{fig:overview}}
	\end{figure*}
	\section{Methodology}
	
	Throughout this paper I will display results obtained by executing code developed for a combination of IBM's Qiskit and Google's quantum AI Cirq. The key ingredient of the code is a method for exactly preparing eigenstates of quadratic Hamiltonians. Preparing eigenstates of quadratic Hamiltonians is equivalent to preparing Slater determinants. Such states are called fermionic Gaussian states, as explained in Ref. \cite{PhysRevApplied.9.044036} and are implemented with Google Quantum AI's Cirq OpenFermion \cite{mcclean2020openfermion}. Fermionic Gaussian states cannot describe excited states of exactly degenerate Hamiltonians \cite{greplova2013quantum}. Such wavefunctions are denoted as $|\psi\rangle$. As extensive numerical testing of the Kitaev chain showed multiple degeneracies at $\mu=0$, I come as close to zero as $\mu=10^{-8} |t|$.
	
	The full $n$-site Kitaev chain Hamiltonian is given by
	
	\begin{equation}\label{eq:Kit}
	H=\sum_{k=1,n}\mu_k c^\dagger_k c _k-\sum_{\langle kj \rangle} \left(t_{kj}c_k^\dagger c_j-\Delta_{kj} c_k^\dagger c_j^\dagger +H.c.\right).
	\end{equation}
	Here, $\mu_k$ denotes the chemical potential at $k$th site, $t_{kj}$ denotes the tunnel hopping between sites $k$ and $j$, $\Delta_{kj}$ the superconducting pairing, $c_k$ annihilates an electron at site $k$ while $c_k^\dagger$ creates the electron at the same site, and H.c. stands for an Hermitian conjugation. Majorana zero modes exist as solutions of the Kitaev Hamiltonian around $\mu=0$ and $\Delta=-t$. This can be seen when substituting ${c^\dagger_k=(\gamma_{2k-1}+i\gamma_{2k})/2}$ and ${c_k=(\gamma_{2k-1}-i\gamma_{2k})/2}$ into Eq. (\ref{eq:Kit}), where $\gamma_k$ is $k$th Majorana operator with the property $\gamma_k=\gamma_k^\dagger$. Throughout this paper, I will assume no local variations of these variables on different sites, hence $\mu_k$, $t_{kj}$ and $\Delta_{kj}$ become $\mu$, $t$ and $\Delta$.
	
	I will calculate and measure expectation values of a number of observables, the expectation value of energy $E=\langle \psi |H| \psi \rangle$, Majorana edge correlation function $\langle \psi|i \gamma_1\gamma_{2n}|\psi \rangle$ \cite{miao2018majorana} (where $n$ denotes the number of sites and $i$ denotes a complex number), Majorana site correlation function $\langle \psi|i \gamma_1\gamma_{k}|\psi \rangle$, Majorana parity operator $\langle\mathcal{P}\rangle=\langle\psi|\prod_{k=1}^{n}\left(1-2c_k^\dagger c_k\right)|\psi\rangle$ and particle number operator $\langle N\rangle=\langle \psi|\sum_{k=1,n}c_k^\dagger c_k|\psi\rangle$. All of these quantities are expressed via fermionic creation and annihilation operators and transformed into a qubit representation via a Jordan-Wigner transformation \cite{tranter2018comparison}.
	
	According to Eq. (17) in Ref. \cite{hegde2015quench}, a topological state is supposed to exhibit parity switches at a chemical potential
	
	\begin{equation}\label{eq:paritys}
	\mu_{\rm PS}=\pm2\sqrt{t^2-\Delta^2}\cos{\left(\frac{\pi p}{n+1}\right)},
	\end{equation}
	where $p=1,...,n/2$ for an even number of sites in the Kitaev chain $n$ and $p=1,...,(n-1)/2$ for an odd $n$.
	
	In FIG. \ref{fig:circ} I display a quantum computer circuit for generating a ground state of a $3$-site Kitaev Hamiltonian at $t=-1$, $\Delta=1$ and $\mu=0^+=10^{-8}$, denoted by $[\,]$. This circuit is composed from nearest-neighbour two-qubit gates defined by $RYXXY(\alpha)=\exp{\left(-i(X\otimes Y-Y\otimes X)\alpha/2\right)}$, and $RZ(\beta)$ gates (rotations of the qubit around the $z$-axis of the Bloch sphere for an angle $\beta$) and $X$ gate rotates a qubit around the $x$-axis of the Bloch sphere for an angle of $\pi$. All excited energy eigenstates of the $3-$site Kitaev Hamiltonian at the given parameter regime are built by first applying $X$ gates to appropriate qubits and then executing the circuit in FIG. \ref{fig:circ}. For instance the first excited state is obtained by applying an $X$ gate to qubit $q0$ followed by the execution of the circuit in FIG. \ref{fig:circ} and is denoted as $[0]$. The highest energy state is obtained by applying $X$ gates to qubits $q0$, $q1$, $q2$ followed by the execution of the circuit in FIG. \ref{fig:circ}, and is denoted as $[0,1,2]$. Google Quantum AI's Cirq calculates the optimal angles $\alpha$ and $\beta$ based on the input of $\mu$, $\Delta$ and $t$. A more detailed discussion on how such circuits are generated is given in Supplementary Material Section S1.\\
	
	\section{A comparison between theory and experiment}
	In FIG. \ref{fig:overview} (a) I display the spectrum of a $3$-site Kitaev chain Hamiltonian having $2^n=8$ eigenstates. Here, a comparison is given between a full diagonalization of the Kitaev Hamiltonian (black "x" symbols) and a solution obtained by implementing Gaussian states on an ideal quantum computer simulator (red and blue full lines). The red (blue) colour denotes states for which $\langle \mathcal{P}(\mu=0^+)\rangle=+1(-1)$. Upon visual comparison these solutions show excellent agreement. The spectrum in subfigure (a) has a lowest energy state (highest energy state) with a next higher (lower) energy state degenerate to it around $\mu=0$, and such degeneracies are marked with arrows. It should be noticed that there is a striking matching between a full numerical diagonalization and the fermionic Gaussian states. 
	
	In subfigure (b) I compare Gaussian states on an ideal simulator of quantum computers with their realisation on IBMQ Santiago. The circuit of the quantum eigenstate of the Kitaev chain is composed out of 6 two-qubit gates ($RYXXY$ is implemented with two CNOT + single-qubit gates) and 9-13 single-qubit gates. The experiment is conducted for 8192 shots, with CNOT errors of $0.74\%$ and readout errors of $1.5\%$. Single-qubit errors are not specified, however the average single-qubit frequency is $4.7$ GHz with a pure dephasing time $T_2=70$ $\mu$s and qubit relaxation time $T_1=83$ $\mu$s. Consequently, single-qubit gate errors influence the results much less than readout and two-qubit gate infidelity. Although all eigenstates move towards zero energy due to quantum noise, the degeneracy between states $[\,]$ and $[0]$ is not lifted by quantum noise. Similarly to that, the degeneracy between states $[1,2]$ and $[0,1,2]$ follows the same pattern. One possible explanation for this degeneracy would be that it is topological in nature which would be the case if states $[0]$ and $[1,2]$ are indeed Majorana zero modes.
	
	The BdG Hamiltonian solves the problem of the Kitaev chain in a single-particle picture. It features electrons and holes and their energy splitting from their ground state (that of electrons and that of holes). Even though the Kitaev chain is rather short, the BdG Hamiltonian (subfigure (c)) is indicating the presence of two zero energy eigenstates (zero in the context of how far away are they from their respective groundstate) which split in energy as the chemical potential $\mu$ is varied. This robust feature exist both in theory and on a noisy quantum computer. The BdG Hamiltonian has $2n=6$ eigenstates for a $3$-site Kitaev chain. For the alleged MZM states, the BdG energy is in remarkable visual accordance with the measured output of the quantum device. One can quantify this by defining a mean absolute error as ${\rm ME}=\left(\sum_i |x_i-y_i|\right)/m$, where $m$ is the total number of measurements/predictions, $x_i$ is the $i$th measurement of energy from the quantum device and $y_i$ is the $i$th prediction of the BdG Hamiltonian. I find a ${\rm ME}=0.129$ for $-1.87 \le y_i\le 1.87$. 
	To further corroborate the correlation between the model and the data I have performed an estimation of the $R^2$ parameter with scipy.stats.linregress and obtained $0.95$ $[0,1,2]-[1,2]$ and $0.75$ for $[]-[0]$ indicating a strong correlation between the model and measured data, even in the presence of quantum noise.
	
	When $t=-\Delta$ an n-site Kitaev chain remains in the topological regime up to $\mu=2t(1-1/n)$. In the case of $n=3$ the system is in the topological regime up to $\mu=4/3$ - first 9 points in the subfigure (c) are topological. For an in-depth discussion on how this condition is calculated the reader is refereed to Supplementary Material Section S2. States which are split from zero energy at low $\mu$ are often referred to as topologically trivial states in literature. It should be noted that although Majorana zero modes remained at zero energy, the topologically trivial states are further shifted towards zero as compared to theory due to quantum noise at low $\mu$. For a more quantitative analysis of the noise present in the system I revert the reader to Supplementary Material S3. 
	
	To further corroborate the Majorana zero mode nature of eigenstates of the Kitaev Hamiltonian I performed experiments and calculated the Majorana edge correlation function, $\langle i \gamma_1 \gamma_6 \rangle$, where $1$ and $6$ are indices of the Majoranas on the edge of the Kitaev chain. If this number is $+1(-1)$ this would mean that the Majoranas on the edges are correlated(anti-correlated) and if this number is $0$ there is no Majorana pairing in the system. In subfigure (d) we see that the states $[0]$ and $[1,2]$ behave exactly as predicted by mean-field theory, as the chemical potential is varied the Majorana states at the edges of the chain become less (anti)-correlated. However, the experimental value does not reach $\pm1$ due to quantum noise. For a more quantitative analysis of the Majorana edge correlation function I revert the reader to Supplementary Material S4.
	
	\begin{figure*}
		\begin{minipage}{0.33\textwidth}
			\subfloat{(a)}{	
				
				\includegraphics[height=3.9cm]{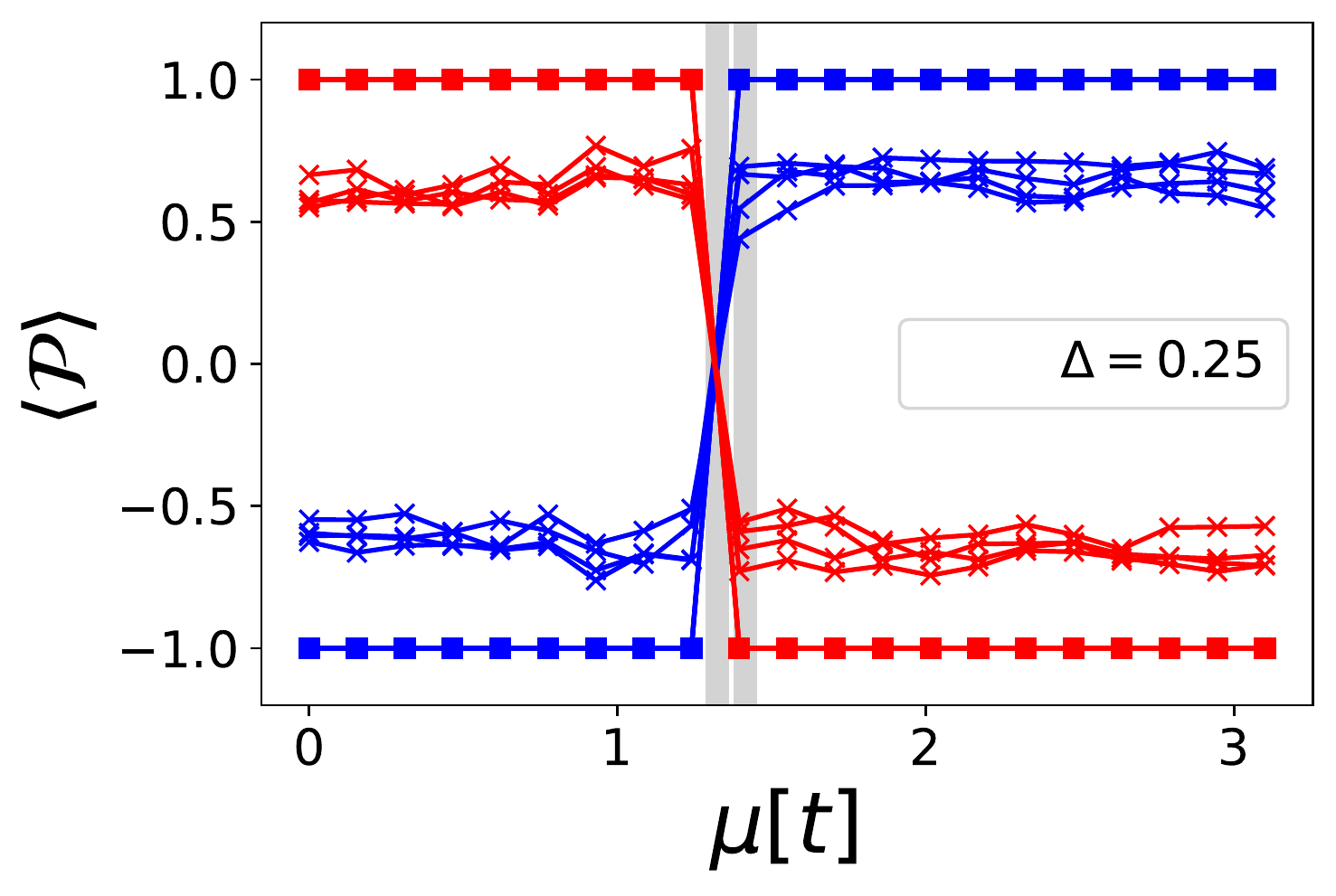}}
		\end{minipage}\begin{minipage}{0.33\textwidth}
			\subfloat{(b)}{	
				
				\includegraphics[height=3.9cm]{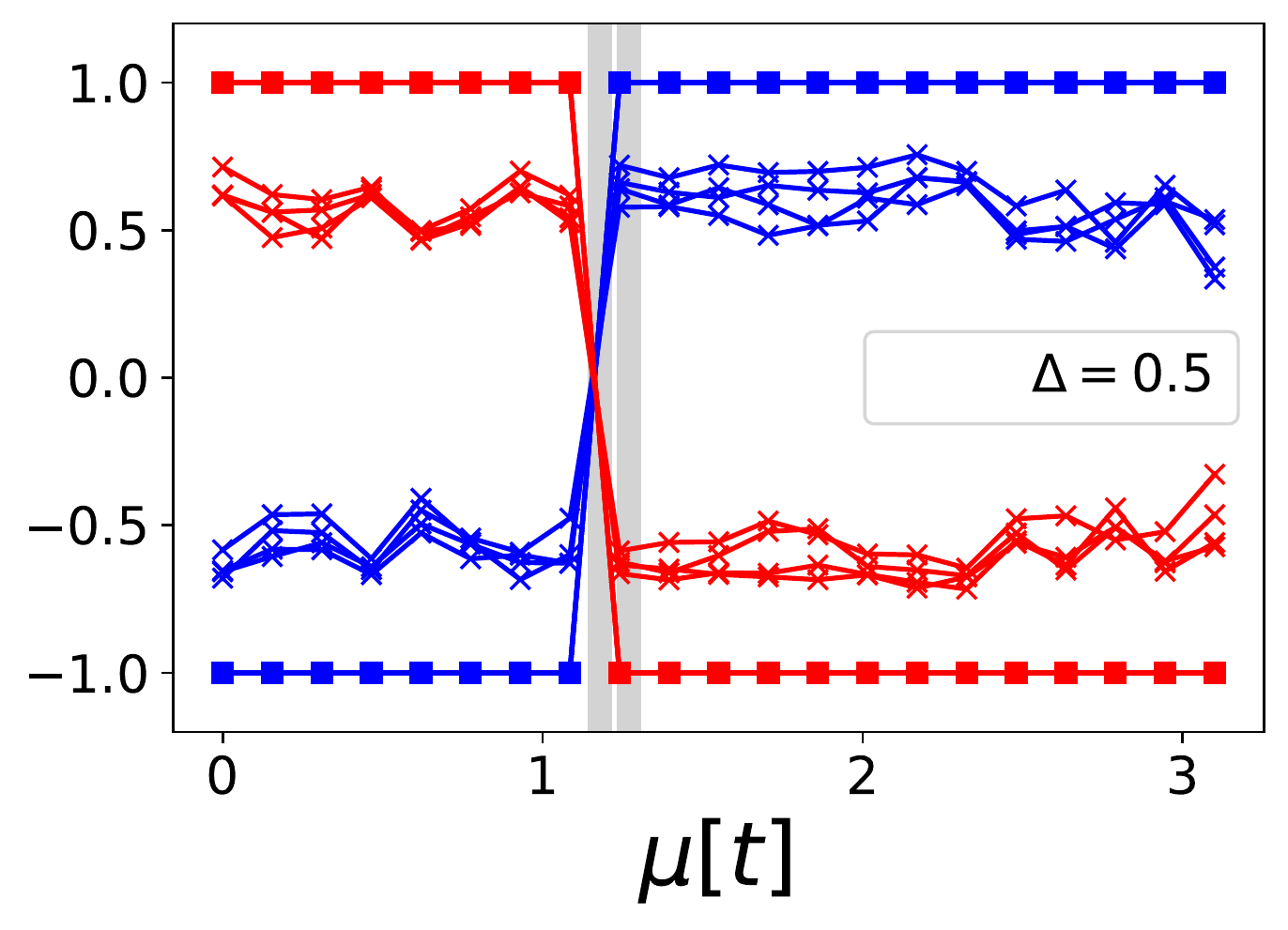}}
		\end{minipage}\begin{minipage}{0.33\textwidth}
			\subfloat{(c)}{	 
				
				\includegraphics[height=3.9cm]{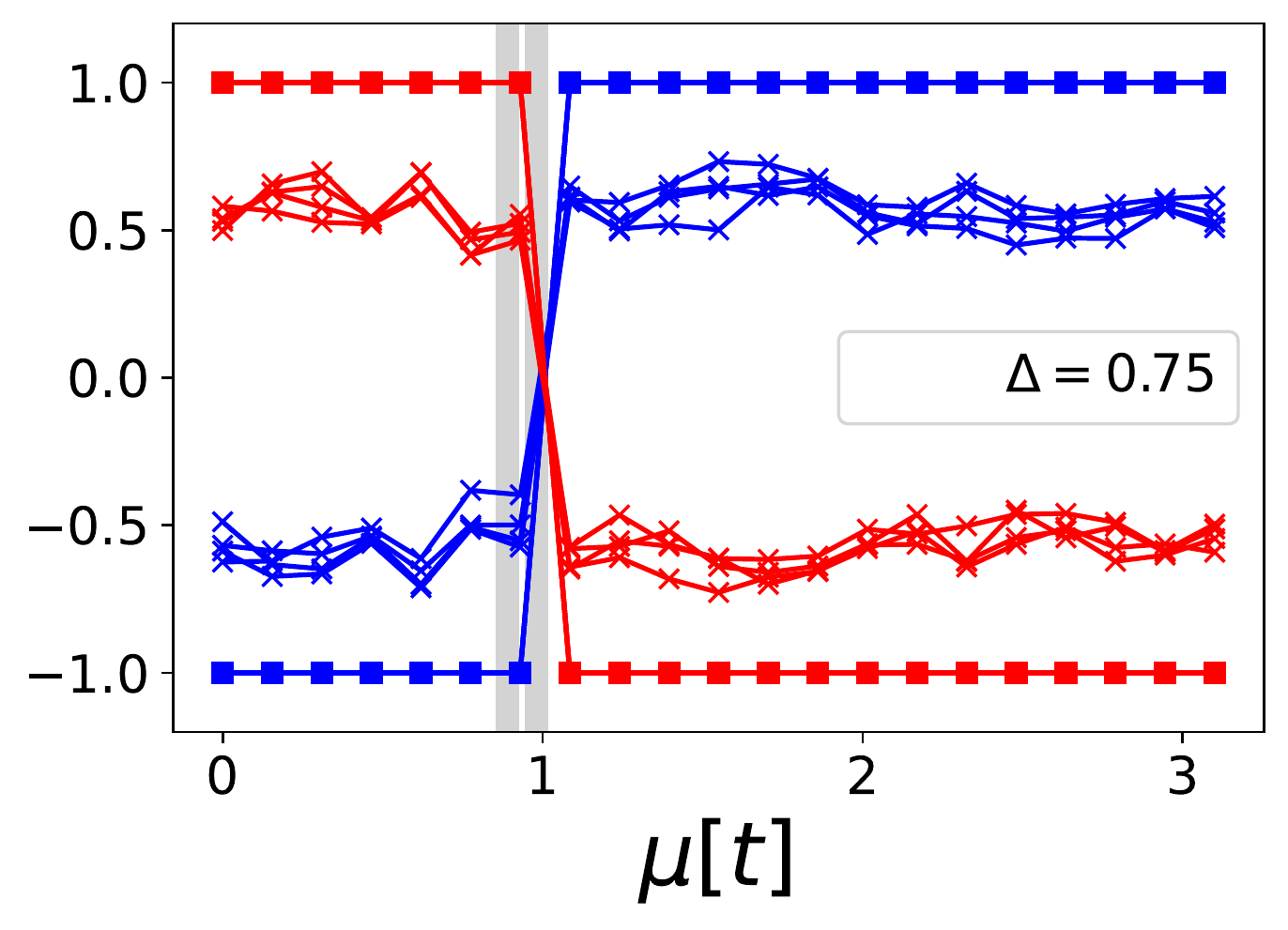}}
		\end{minipage}
		
		\begin{minipage}{0.33\textwidth}
			\subfloat{(d)}{	
				
				\includegraphics[height=3.9cm]{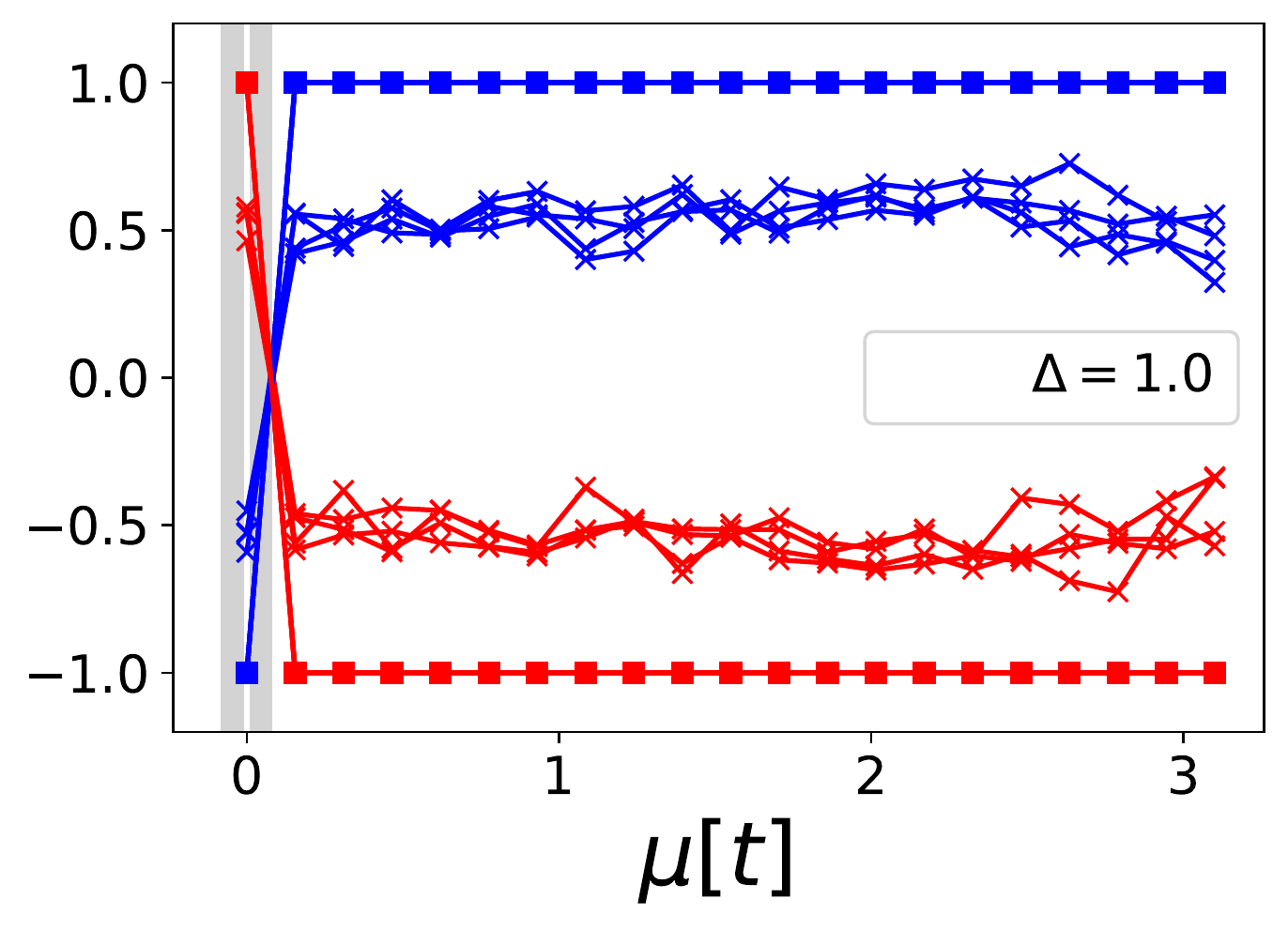}}
		\end{minipage}\begin{minipage}{0.33\textwidth}
			\subfloat{(e)}{	
				
				\includegraphics[height=3.9cm]{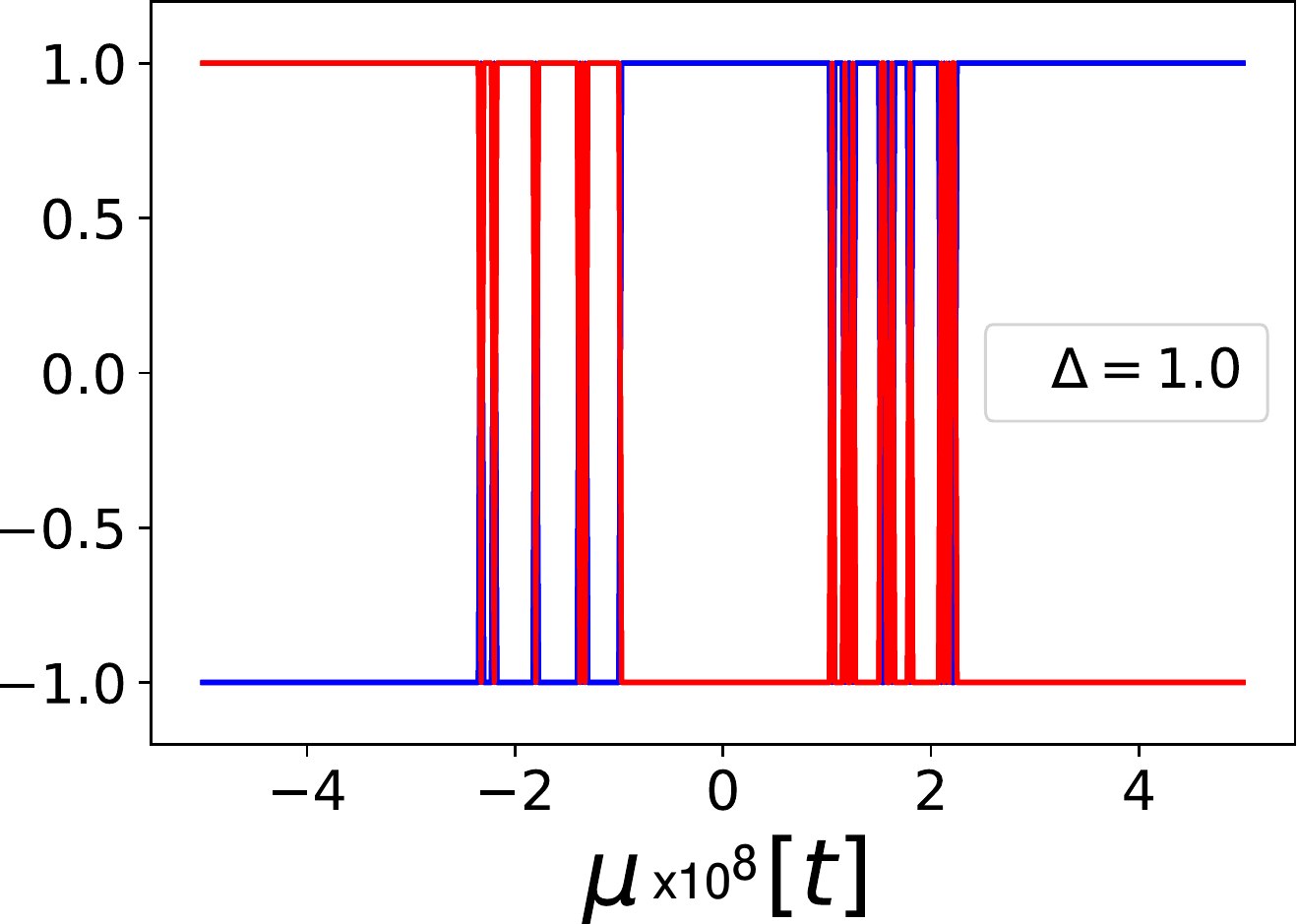}}
		\end{minipage}\begin{minipage}{0.33\textwidth}
			\subfloat{(f)}{	
				
				\includegraphics[height=3.9cm]{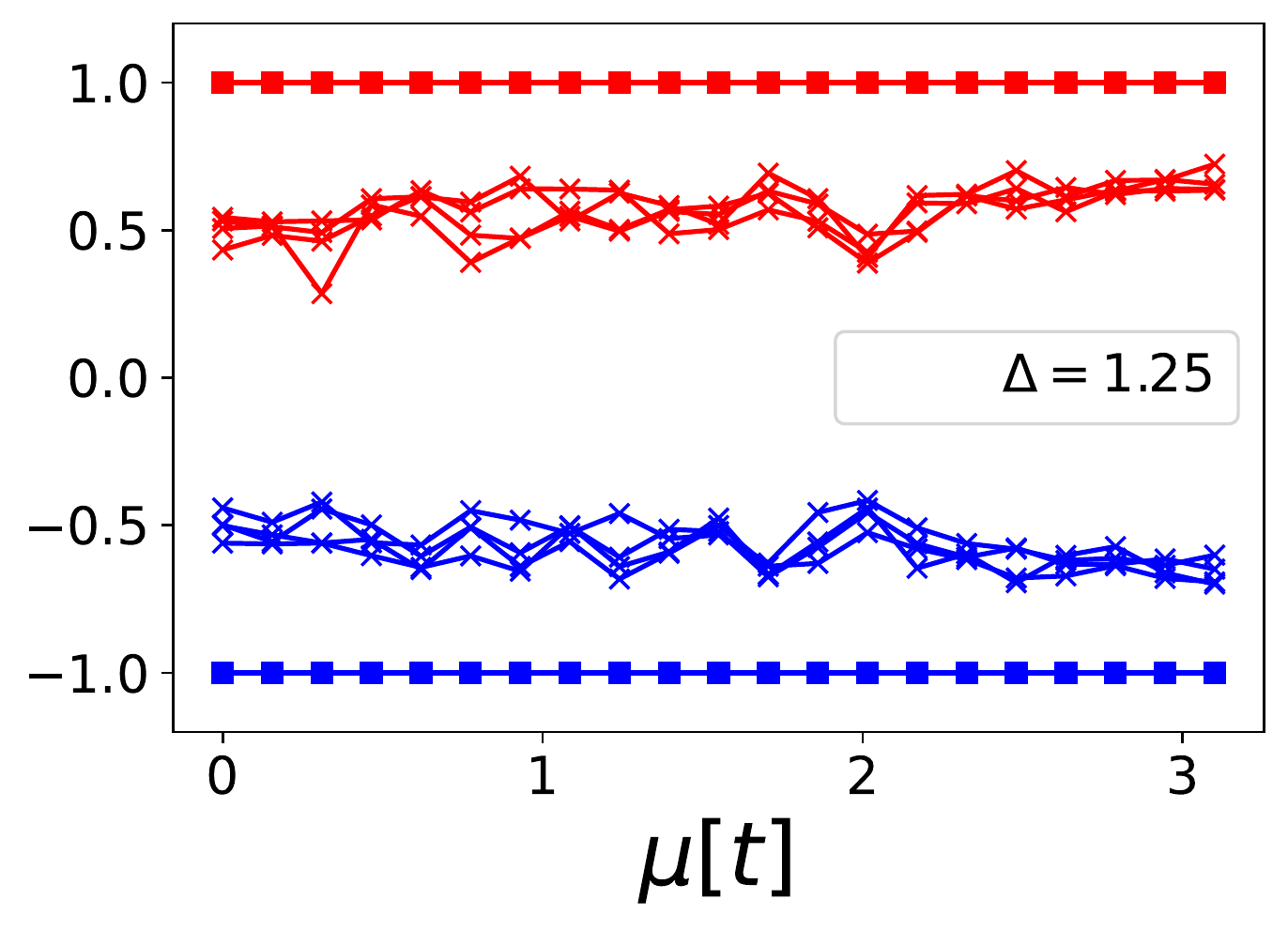}}	
		\end{minipage}	
		
		\caption{(a-d) and (f) Parity switches at $t=-1$ for different values of $\Delta$ and for 20 values of $\mu$ between $10^{-8}$ and $3.1$ with a $0.155$ increment. (e) Parity switches for 400 values of $\mu$ between $-5\cdot 10^{-8}$ to $5\cdot 10^{-8}$. \label{fig:paritys}}. 
	\end{figure*}
	
	As further consistency checks, I performed measurements and simulations of the parity of the eigenstates of the Kitaev Hamiltonian, and these results are displayed in subfigure (e). The eigenstates are clearly separated in terms of parity $\pm 1$ in case of theory, and positive and negative in case of experiment. A naive expectation for MZMs is that they do conserve parity as pairs of particles can be freely generated from the superconducting condensate. Similarly to energy measurements, the discrepancy between the theoretical and the experimental values is due to quantum noise. It should be noted that the eigenstates undergo a topological phase transition in which the parity of all eigenstates switches between the first two points ($\mu_1=10^{-8}$ and $\mu_2=0.155$). This topological phase transition is predicted to occur within the single-particle picture at $\mu=0$ and will be discussed in detail later in this section.
	
	Lastly in subfigure (f) I plot the total particle number of particles for different eigenstates. A naive expectation for MZMs is that they do not conserve particle number as a function of chemical potential, at least in a picture where the topological region is observed separately from the bare superconductor \cite{lin2018towards, PhysRevLett.124.257002}. This is due to the fact that in the picture of a separated topological and non-topological condensate pairs of particles can be freely generated from the superconducting condensate \cite{lin2018towards}, and states $[0]$ and $[1,2]$ behave in the expected manner. If particle number was conserved (a good quantum number) this would mean that as the parameters of the Hamiltonian and respective eigenstate are changed, the expectation value of the particle number operator would remain the same. For states $[0]$ and $[1,2]$ the particle number is not conserved at small $\mu$ and then saturates to a value $1$ or $2$ at large values of $\mu$.
	
	In FIG. \ref{fig:paritys} I test the single-particle prediction that parity switches occur according to Eq. (\ref{eq:paritys}). Tunnelling is kept constant at $t=-1$, $\Delta=0.25,0.5,0.75,1,1.25$ and $\mu$ is varied between $10^{-8}$ and $3.1$ in 20 increments of $0.155$. The points where parity switches represent a topological phase transition. Squared markers are added to ideal quantum simulator data. The grey line represents a region where the parity switching is likely to occur and it exists due to a limited resolution in $\mu$ in which the experiment is performed. The white line is the exact expected position of such a transition as predicted by the single-particle picture. Such parity switches could be understood as points in which two MZM eigenstates cross the zero of energy.
	
	\begin{figure*}[t!]
		\begin{minipage}{0.38\textwidth}
			\centering  
			\subfloat{(a) $\mu=10^{-8}$}{

				\includegraphics[height=3.7cm]{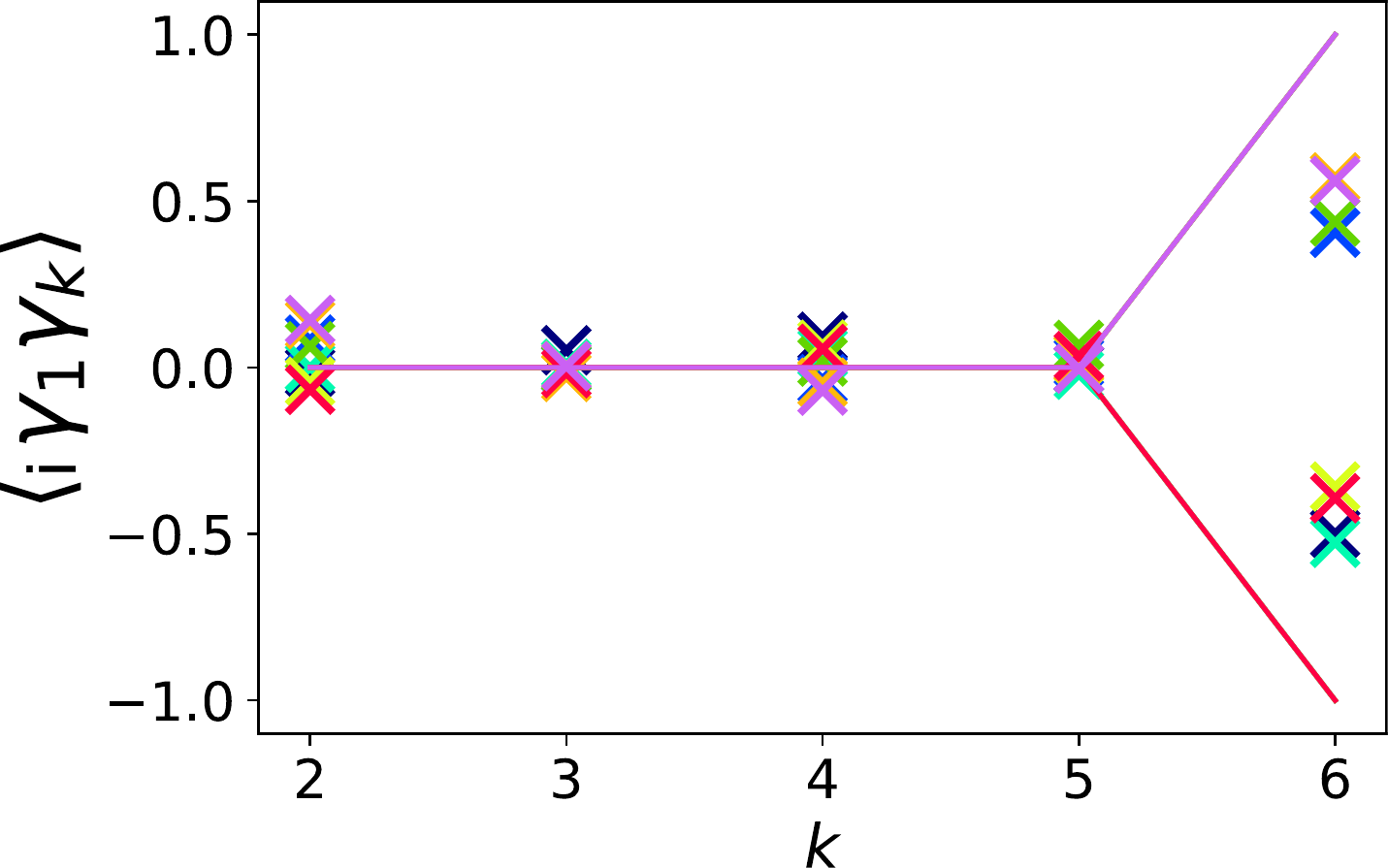}}
		\end{minipage}\begin{minipage}{0.3\textwidth}
			\subfloat{(b) $\mu=1$}{	
				
				\includegraphics[height=3.7cm]{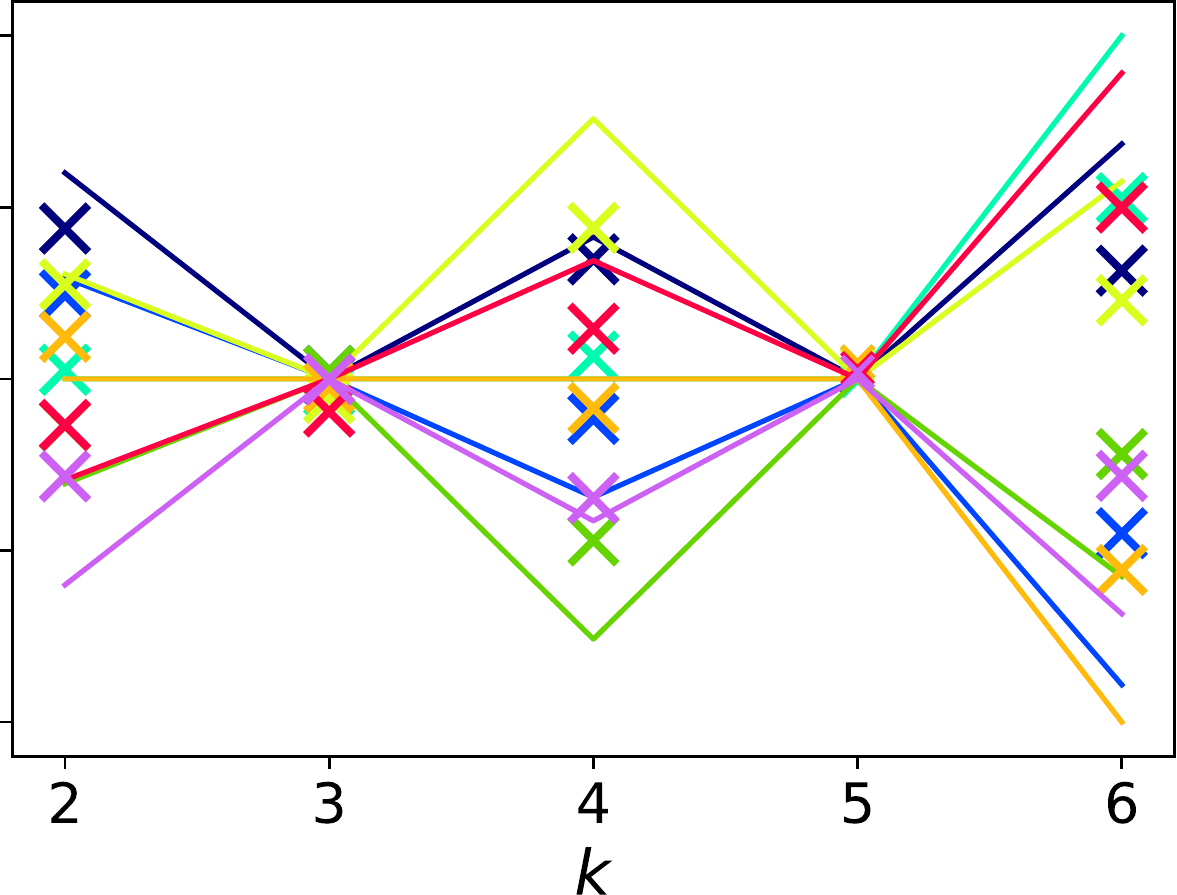}}
		\end{minipage}\begin{minipage}{0.33\textwidth}	
			\subfloat{(c) $\mu=2$}{	
				
				\includegraphics[height=3.7cm]{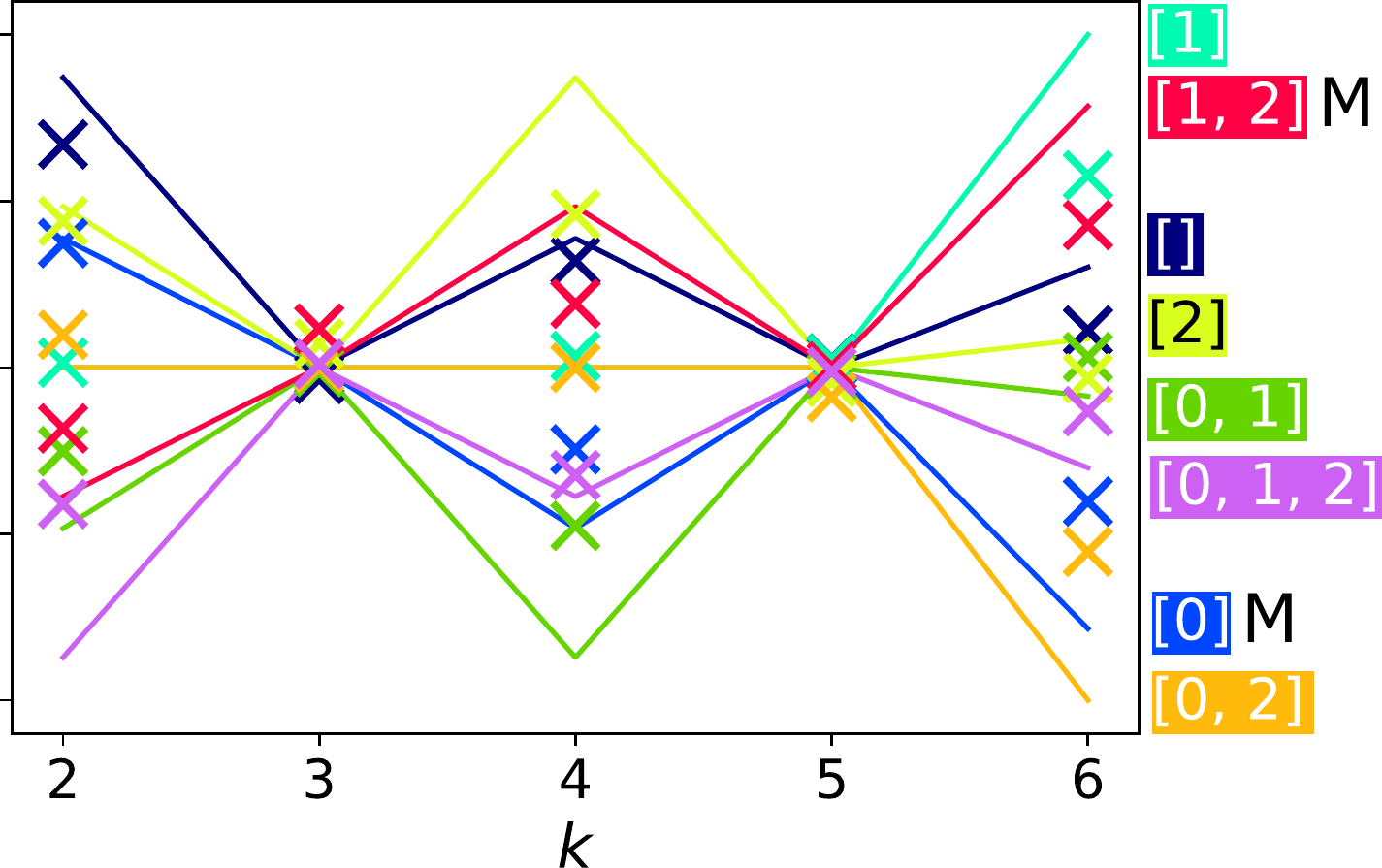}}
			
		\end{minipage}
		\caption{Site correlation function where $k$ denotes the site at different values of the chemical potential in the topological regime $\Delta=-t=1$. Full lines denote values obtained by an ideal simulator of quantum computers at points of integer $k$ and "x"-markers are output from IBMQ Santiago. Colour code is explained on the right.\label{fig:corr}}
	\end{figure*}
	
	The results show that the behaviour of the exact solutions and the experiment is much in line with the prediction: parity switches occur exactly where they are predicted with Eq. (\ref{eq:paritys}). Here, I find the only minor discrepancy between the single-particle picture and the full solution - parity switches occur in a narrow parameter region in $\mu$ (see subfigure (e)) and not in a single point. However, this region in parameter space of $\mu$ is quite narrow - on the order of $4\cdot 10^{-8}$ that I conclude that it is quite in line with predictions. A comparison between the simulated results (squares) and outputs of the quantum computers ("x" markers) show that this feature is quite robust to quantum noise. Although the value of the parity decreases on quantum computers as compared to ideal simulations, the point where parity switches is robust to any relaxation and pure dephasing. This an indication that this parity switches represent a topological phase transition.

	In FIG. \ref{fig:corr} we observe the Majorana site correlation function $\langle i \gamma_1 \gamma_k \rangle$ in the topological regime $\Delta=-t=1$ as a function of the chemical potential $\mu$. First it should be noted that the operator $i \gamma_1 \gamma_k $ is non-Hermitian at $k=1$ so the site correlation operator is an observable only when $k>1$. 
	
	At low chemical potential states $[0]$ and $[1,2]$ are exhibiting a Majorana-like character - a Majorana site correlation function localized at edges which is $\pm 1$ for the results of an ideal simulation (full lines). The actual execution on a quantum computer (colored "x" symbols) follows a similar qualitative trend but does not quantitatively reach $\pm 1$ due to quantum noise. As the chemical potential increased, Majorana zero modes start favoring correlations between neighboring Majorana fermions more. This is potentially a key differentiation between MZMs and trivial zero-energy states such as Andreev bound states, as the latter are not localized at the edges \cite{prada2020andreev}.
	
\begin{figure}[t!]
		\includegraphics[height=4.5cm]{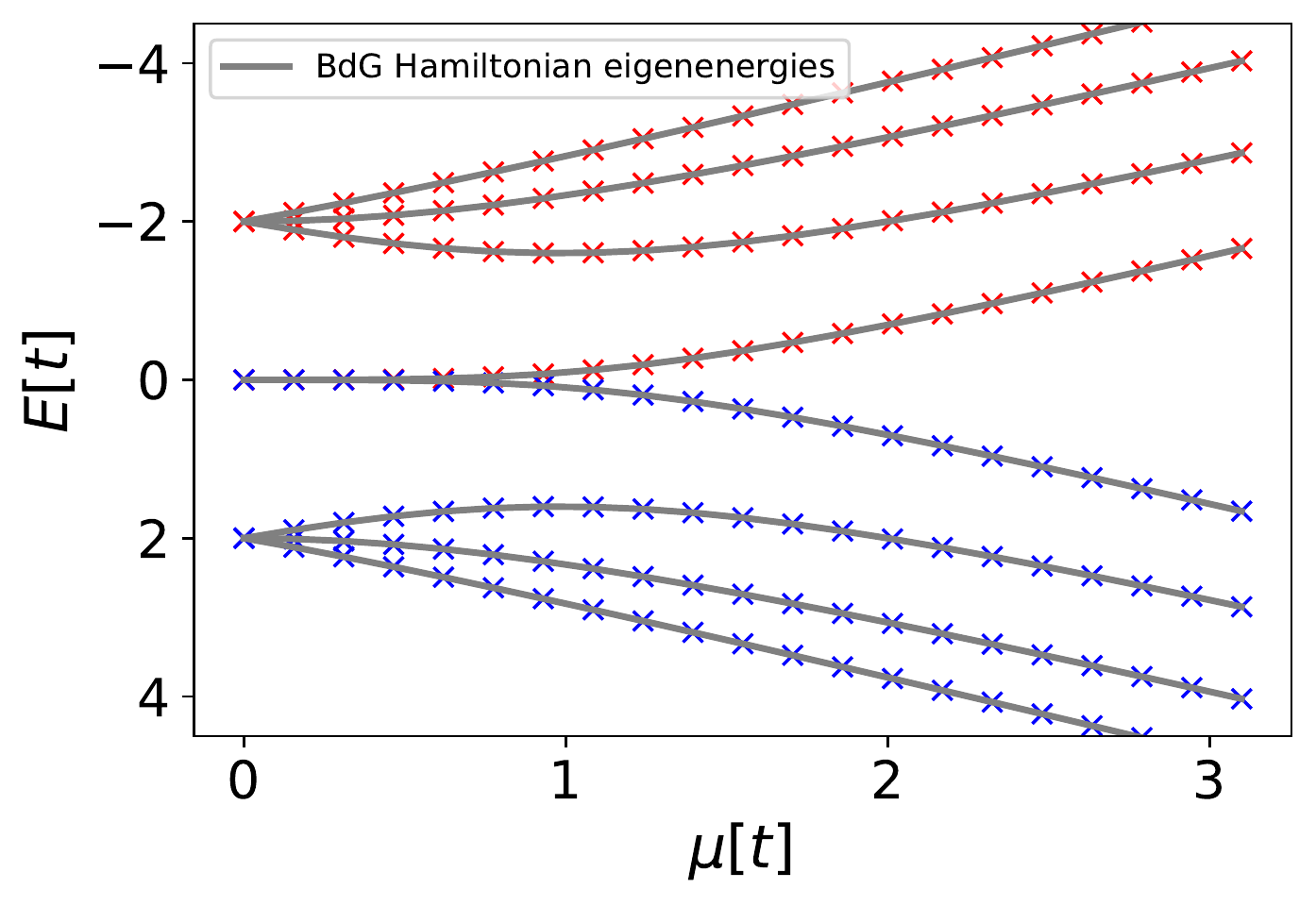}
		\caption{Quantities of a $4$-site Kitaev Hamiltonian at $t=-1$ and $\Delta=1$ as a function of the chemical potential $\mu$ in units of absolute value of tunnelling $[t]$. Red denotes values obtained on an ideal simulator of quantum computers with $\langle P(\mu=0^+)\rangle=1$. Blue denotes values obtained on an ideal simulator of quantum computers with $\langle P(\mu=0^+)\rangle=-1$. The single-particle picture BdG spectrum, diagonalizatioin of the BdG Hamiltonian (grey lines) compared to Gaussian state solutions (coloured markers). \label{fig:4site_main}}
	\end{figure}

	This figure also allows the determination of the type of quantum noise dominating in the experiment. When observing the value of the site-correlation function for states $[\,]$ and $[0,1,2]$ at $\mu=1$ and $\mu=2$ around $k=2$ we see that the state $[0,1,2]$ and the state $[\,]$ have the same absolute value in theory. However, in the experimental realization the measured value of the site correlation function is much closer to the theoretical value for the state $[\,]$ as opposed to the state $[0,1,2]$. The only difference between these two states are the simulations three single qubit $X$ gates applied to qubits $q0$, $q1$ and $q2$. It should be noted that single qubit gates by themselves should not have such a profound effect on state fidelity due to the fact that they are quite noise robust. This is a strong indication that qubit cross-talk is present and the dominating dephasing source in IBMQ Santiago -  an effect already known and well characterized for other IBMQ processors such as IBMQ Poughkeepsie \cite{murali2020software}. In Supplementary Material S5 I give another realization of the same experiment corroborating the same qualitative features. In Supplementary material S6 the same results are presented for a 4-site Kitaev chain on simulators of quantum computers. Example of a 4-site BdG spectrum is given here just to show that the methodology generalizes to longer chains Fig. \ref{fig:4site_main}
	
	\textit{Comparison with non-topological phenomena mimicking MZMs -} A number of topologically trivial phenomena mimic the behavior of MZMs in some regards.  Caroli-de Gennes-Matricon (CdGM) states \cite{chen2018discrete}, Yu-Shiba-Rusinov (YSR) states \cite{RevModPhys.78.373} and Andreev bound states \cite{prada2020andreev} are the most common ones refereed to in the literature. CdGM states are subgap states close to zero energy when $\Delta \ll \mu$. In our case the sticking of the energy levels to zero occurs in the opposite limit when $\Delta \gg \mu$. Furthermore, YSR states and Andreev bound states require a non-superconducting region. A bound state which is created in these two cases is local in nature \cite{prada2020andreev}- unlike the case presented here no long-range Majorana correlations are present.

	\section{Conclusion} To conclude, here I solved the Kitaev chain exactly with quantum computing methods and showed both theoretically and experimentally that two eigenstates of the Kitaev Hamiltonian have a large number of features that corroborate their Majorana zero mode nature. They are zero energy excitations of their respective groundstate with a robust-to-noise degeneracy, they have a Majorana edge-correlation function which decays with the chemical potential. Furthermore, Majorana zero modes favour Majorana pairing between edges of the Kitaev chain, do not preserve particle number and have parity switches at points in the parameter space as predicted by single-particle theories. The results presented here are the most complete set of experimental validations which confirm the existence of Majorana zero modes on the edges of a Kitaev chain, as the eigenstates of the Kitaev Hamiltonian are tested for eight distinct indications of MZMs in a single, reproducible experiment.
	
	\section{Acknowledgments}
	I would like to thank endlessly A. Lj. Ran\v{c}i\'{c} for all her patience, support and science discussions during the course of this work. I am also grateful to O. Dmytruk for live discussions and explanations.
	
	\section{Data availability} 
	Full experimental data and code is available at \href{https://zenodo.org/record/6323467
	}{10.5281/zenodo.6323467}.
	
	\section{References}
	\bibliography{PaperV_1.bib}

\clearpage

\newcommand\myeq{\stackrel{\mathclap{\normalfont\mbox{\tiny n=3}}}{=}}
\newcommand{\highlight}[1]{\colorbox{yellow}{$\displaystyle #1$}}

\begin{center}
	\textbf{\large Supplementary Material: Exactly solving the Kitaev chain and generating Majorana-zero-modes out of noisy qubits}
\end{center}

	\author{Marko J. Ran\v{c}i\'{c}}
	
	\email{marko.rancic@totalenergies.com}
	
	\address{Address TotalEnergies, Tour Coupole La Défense, 2 Pl. Jean Millier, 92078 Paris}
	
	\setcounter{equation}{0}
	\setcounter{section}{0}
	\setcounter{figure}{0}
	\setcounter{table}{0}
	\setcounter{page}{1}
	\makeatletter
	\renewcommand{\theequation}{S\arabic{equation}}
	\renewcommand{\thefigure}{S\arabic{figure}}
	\renewcommand{\thesection}{S\arabic{section}}
	
	\maketitle
	\section{S1 - Preparing eigenstates of quadratic Hamiltonians on a quantum computer}
	In this Section I will briefly outline the procedure for obtaining the ground state of quadratic Hamiltonians relying heavily on steps described in Appendix A and Section IV of Ref. \cite{App9044036}. I will try to use the same notation and follow exactly the same steps. The reader is refereed to the original study for a more detailed discussion. This procedure is implemented in Google's Quantum AI Cirq in an exact numerical fashion \cite{mcclean2020openfermion}. Afterwards, The circuit obtained in Google's Quantum AI Cirq is converted to a Qiskit form and the code for that is given in \href{https://zenodo.org/record/6323467
	}{10.5281/zenodo.6323467}.
	
	The procedure is as follows:
	
	1. Find $A$ the Majorana representation of a general quadratic Hamiltonian $\mathcal{H}=\frac{i}{2}\mathbf{f}^T A \mathbf{f}$.
	
	2. Bring the matrix  $A$ into the Schur form and permute indices such that the upper block $\varepsilon$ is diagonal with eigenvalues in increasing order
	
	\begin{equation}
	{\rm perm}({\rm Schur}(A))=R A R^T=
	\begin{bmatrix}
	0  & \varepsilon\\
	-\varepsilon & 0\\
	\end{bmatrix}.
	\end{equation}
	
	3. Obtain $W=\Omega^\dagger R \Omega$, where
	
	\begin{equation}
	\Omega=\frac{1}{2}\begin{pmatrix}
	\mathbbm{1}&& \mathbbm{1}\\
	i\mathbbm{1}&&-i\mathbbm{1}
	\end{pmatrix}.
	\end{equation}
	The form of $W$ is in general
	
	\begin{equation}
	W=\begin{bmatrix}
	W_1^* && W_2^*\\
	W_2 && W_1
	\end{bmatrix}.
	\end{equation} 
	
	4. Find such unitary transformation $U$ for which $VW_LU=[0 \,\mathbbm{1}]$, where $V$ is a general unitary matrix, $W_L=[W_1^* W_2^*]$. 
	
	5. $U$ can be decomposed into products of $X$ gates on the last qubit and a combination of $RYXXY$ and single qubit $Z$ gates - and it is a representation of a unitary operator which diagonalizes a general quadratic Hamiltonian.
	
	\section{S2 - The winding number - a topological invariant for a finite Kitaev chain}
	
	In this Section I will give an answer to the question of the boundary of the topological phase of finite Kiteav chains. To paraphrase it more specifically and in a more simple manner: when $t=-\Delta$ and $n$ is finite, for which value of $\mu$ does a topological-non-topological transition occur in the Kitaev Hamiltonian?
	
	The starting point is the derivation is $n$-site Kitaev Hamiltonian
	
	\begin{equation}\label{eq:Kit_sup}
	H=\sum_{k=1,n}\mu_k c^\dagger_k c _k-\sum_{\langle kj \rangle} \left(t_{kj}c_k^\dagger c_j-\Delta_{kj} c_k^\dagger c_j^\dagger +H.c.\right).
	\end{equation}
	Now I perform a discrete Fourier transform of Eq. (\ref{eq:Kit_sup}) by setting $c_k^\dagger=\frac{1}{\sqrt{n}}\sum_K \exp{\left(-ikK\right)}c_K^\dagger$ and $c_k=\frac{1}{\sqrt{n}}\sum_K \exp{\left(ikK\right)}c_K$. Here the capital $K$ denotes the wavenumber while the lowercase $k$ denotes the site in the Kitaev chain. Setting $\mu_k=\mu$, $t_{jk}=t$ and $\Delta_{jk}=\Delta$ the following Hamiltonian is obtained
	
	\begin{multline}\label{eq:Kit_mom}
	H=-\sum_{K}c_K^\dagger c_K\left(\mu+2t\left(1-\frac{1}{n}\right)\right)\cos{(K)}+\\
	\Delta \left(1-\frac{1}{n}\right)\sum_K \left(e^{iK} c_{-K}c_{K}+e^{-iK} c^{\dagger}_Kc^{\dagger}_{-K}\right).
	\end{multline}
	The entire procedure between Eq. (\ref{eq:Kit_sup}) to Eq. (\ref{eq:Kit_mom}) can be found for instance in Appendix B of Ref. \cite{skantzaris2014conductivity}. When $n\rightarrow \infty$ the term $1-1/n=1$ leading to the following Hamiltonian in $K$ space
	
	\begin{multline}\label{eq:Kit_mom2}
	H=-\sum_{K}c_K^\dagger c_K\left(\mu+2t\right)\cos{(K)}+\\
	\Delta \sum_K \left(e^{iK} c_{-KK}+e^{-iK} c^{\dagger}_Kc^{\dagger}_{-K}\right).
	\end{multline}
	Going back to finite Kitaev chains and setting $t\left(1-1/n\right)=\tilde{t}$ and $ \Delta\left(1-1/n\right)=\tilde{\Delta}$ in Eq. (\ref{eq:Kit_mom}) the same form of the Hamiltonian as in Eq. ({\ref{eq:Kit_mom2}}) is obtained
	
	\begin{multline}\label{eq:Kit_mom3}
	H=-\sum_{K}c_K^\dagger c_K\left(\mu+2\tilde{t}\right)\cos{(K)}+\\
	\tilde{\Delta} \sum_K \left(e^{iK} c_{-K}c_{K}+e^{-iK} c^{\dagger}_Kc^{\dagger}_{-K}\right).
	\end{multline}
	One can draw a conclusion that the behavior of the infinite Kitaev chains in $K$-space is the same as those of finite chains with a modified tunneling and superconducting pairing $t\rightarrow \tilde{t}$ and $\Delta \rightarrow \tilde{\Delta}$. When $\Delta=-t$ the topological condition in infinite Kitaev chains is $\mu< 2t$ \cite{kitaev2001unpaired,leumer2020exact}. Given that topological properties are fully determined from properties of the $K$-space Hamiltonians \cite{RevModPhys.88.035005} the topological condition for a finite Kitaev chain is 
	
	\begin{equation}\label{eq:cond_1}
	\mu< 2\tilde{t}=2t\left(1-\frac{1}{n}\right) \; \myeq \; \frac{4}{3}t.
	\end{equation}
	
\begin{figure}
		\includegraphics[width=0.5\textwidth]{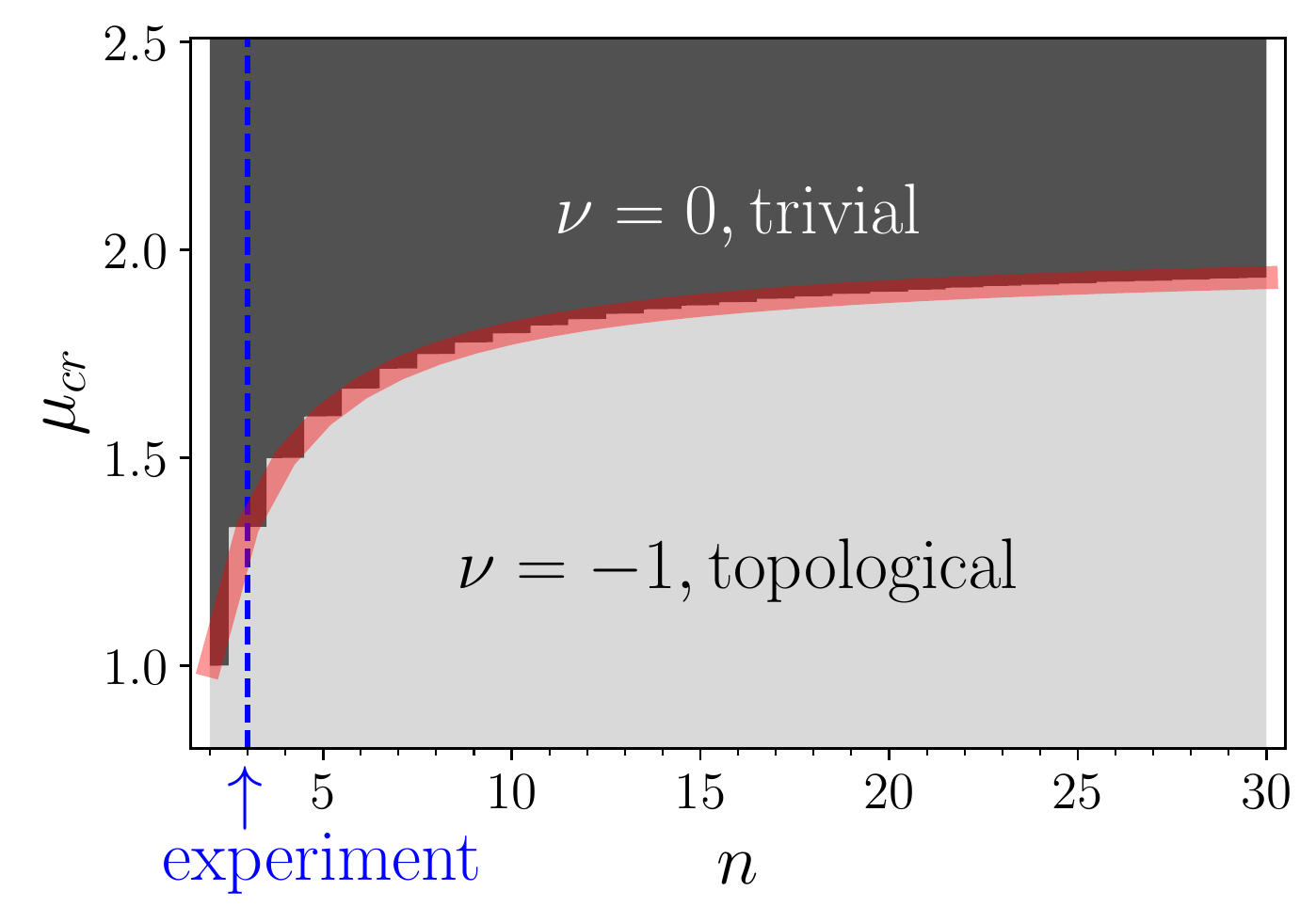}
		\caption{The topological and non-topological region boundary as a function of Kitaev chain length. The figure shows the critical chemical potential which is the boundary between the two phases $\mu_{\rm\textit{cr}}$ when $t=-\Delta=1$. The red line is the condition in Eq. (\ref{eq:cond_1}), while the grey areas are obtained by numerically unwinding and integrating Eq. (\ref{eq:wind}). \label{fig:wind_nr}}
	\end{figure}

	The potentially topological state of matter in a finite Kitaev chain can be further described by a winding number $\nu$ \cite{RevModPhys.88.035005}. Similarly with a Chern number the winding number is a topological invariant of a given system \cite{RevModPhys.88.035005,leumer2020exact}. In the case of the Kitaev chain, $\nu=\pm 1$ would indicate a topological phase and $\nu=0$ a topologically trivial phase. The winding number is defined as \cite{RevModPhys.88.035005,leumer2020exact}
	
	\begin{equation}\label{eq:wind}
	\nu=\frac{1}{2\pi}\int_{-\pi}^{\pi} dK \partial_K w(K),
	\end{equation}
	where $\partial_K w(K)$ is the winding number density and ${w(K)=\rm{Arg} [2\tilde{\Delta} \sin{(K)}+i(\mu+ 2\tilde{t}\cos{(K)})]}$. Here $\rm{Arg}$ stands for a non-principal value of the argument. The integral in Eq. (\ref{eq:wind}) is difficult to compute due to the fact that the non-principal value of the argument cannot be found in an analytical fashion. Still, a numerical routine can be developed, by unwinding an array of different values of $w(K)$ in the region $[-\pi,\pi]$. In the results given at Fig. \ref{fig:wind_nr} we compare results of Eq. (\ref{eq:cond_1}) with a numerical solution of the integral in Eq. (\ref{eq:wind}) obtained by numerically unwinding $w(K)$ in the region $[-\pi,\pi]$ by Python's numpy unwind function.
	
	In Fig. \ref{fig:wind_nr} we see that even the shortest possible Kitaev chain of only 2 sites, remains topological for ${\mu<t=1}$ and that as the length of the chain is increased the topological-non-topological boundary approaches that of infinite chains $\mu<t=2$ \cite{kitaev2001unpaired}. In the  case of the experiment in the main body of the paper $n=3$, $\mu_{\rm\textit{cr}}=4/3$ and so the system remains topological in $75\%$ of values of $\mu$ as compared to an infinite Kitaev chain. This means that the 9 points corresponding to low value of $\mu$ are in the topologically non-trivial regime in the main body of the paper.
	
	Finally, two-qubit gates error rates of contemporary quantum computers ($<0.75\%$) and are the most dominant source of errors on current-day quantum devices. We can try to estimate are such constrains on error rates of two qubit gates preventing one to achieve an MZM state on a quantum computer. 
	
	To measure that we define a parameter which represents a difference between the two-qubit gate angles at $\mu=0^+$ and $\mu\approx 1.33$ normalized to $\pi$ for every two qubit gate in the circuit preparing the ground state of the MZMs (Fig. 1 - main body of the paper)
	
	\begin{equation}
	\delta=\frac{\rm{abs}\left(\theta_{ij}|_{\mu=0+}-\theta_{ij}|_{\mu\approx 1.33}\right)}{\pi}.
	\end{equation}	
	In Tab \ref{tab:tab1} we see that circuit is remaining in the topological regime even if the uncertainty in generating the two qubit gate is on the order of $\sim 0.75\%$ outlining the possibility of creating MZMs on modern gate-based quantum computers. The exception to this is the first two qubit gate between qubits $q_1$ and $q_2$ which has to have a value of $\pm \pi/2$ for a broad range of $\mu$. However, the two-qubit $RYXXY$ gate has a special symmetry point at $\theta_{ij}=\pi/2$ in which it is equivalent to two single qubit gates $Z$-gates which have much lower error rates (on the order or $<0.1\%$)
	
	\begin{equation}
	RYXXY\left(\frac{\pi}{2}\right)=e^{i\frac{\pi}{2}(\sigma_y^1\otimes \sigma_x^2-\sigma_x^1\otimes\sigma_y^2)}=\sigma_z^1\otimes \sigma_z^2.
	\end{equation}
	\begin{table}
		\begin{center}
			\begin{tabular}{ c||c|c|c|c|c|c }
				$\mu$	& $\theta_{12}(\pi)$ & $\theta_{01}(\pi)$ & $\theta_{12}(\pi)$& $\theta_{12}(\pi)$ &  $\theta_{01}(\pi)$ & $\theta_{12}(\pi)$\\
				\hline
				$0^+$ & $0.500$ & $0.304$ & $0.333$ & $0.333$ & $0.304$ & $0.250$\\
				$0.155$ & $0.500$ & $0.317$ & $0.337$ & $0.314$ & $0.301$ & $0.257$\\
				$0.310$ & $0.500$ & $0.327$ & $0.342$ & $0.294$ & $0.298$ & $0.264$\\
				$0.465$ & $0.500$ & $0.334$ & $0.347$ & $0.272$ & $0.294$ & $0.273$\\
				$0.620$ & $-0.500$ & $0.338$ & $0.353$ & $0.249$ & $0.290$ & $0.282$\\
				$0.775$ & $0.500$ & $0.339$ & $0.359$ & $0.227$ & $0.286$ & $0.293$\\
				$0.930$ & $0.500$ & $0.339$ & $0.366$ & $0.205$ & $0.282$ & $0.303$\\
				$1.085$ & $0.500$ & $0.337$ & $0.372$ & $0.186$ & $0.270$ & $0.313$\\
				$1.240$ & $0.500$ & $0.334$ & $0.379$ & $0.169$ & $0.275$ & $0.324$\\
				\hline
				\highlight{$1.395$} & \highlight{$0.500$} & \highlight{$0.331$} & \highlight{$0.385$} & \highlight{$0.154$} & \highlight{$0.273$} & \highlight{$0.334$}\\
				\hline
				\hline
				$\delta$& $0.0$ & $0.9\%$ & $1.5\%$ & $5.2\%$ & $0.9\%$ & $2.4\%$
				
			\end{tabular}
		\end{center}
		\caption{Angles of two qubit $RYXXY$ gates required to prepare Majorana zero modes - no color topological vs. yellow trivial. $\delta$ absolute difference between the two-qubit gates angle of the initial topological and trivial regions. The circuit has the same form like in the main body of the paper in Fig. 1. \label{tab:tab1}}
	\end{table}

	\section{S3 - A noise model}
	
	\begin{figure*}
		\centering
		\begin{minipage}{0.25\textwidth}
			\centering
			\includegraphics[width=1.0\textwidth]{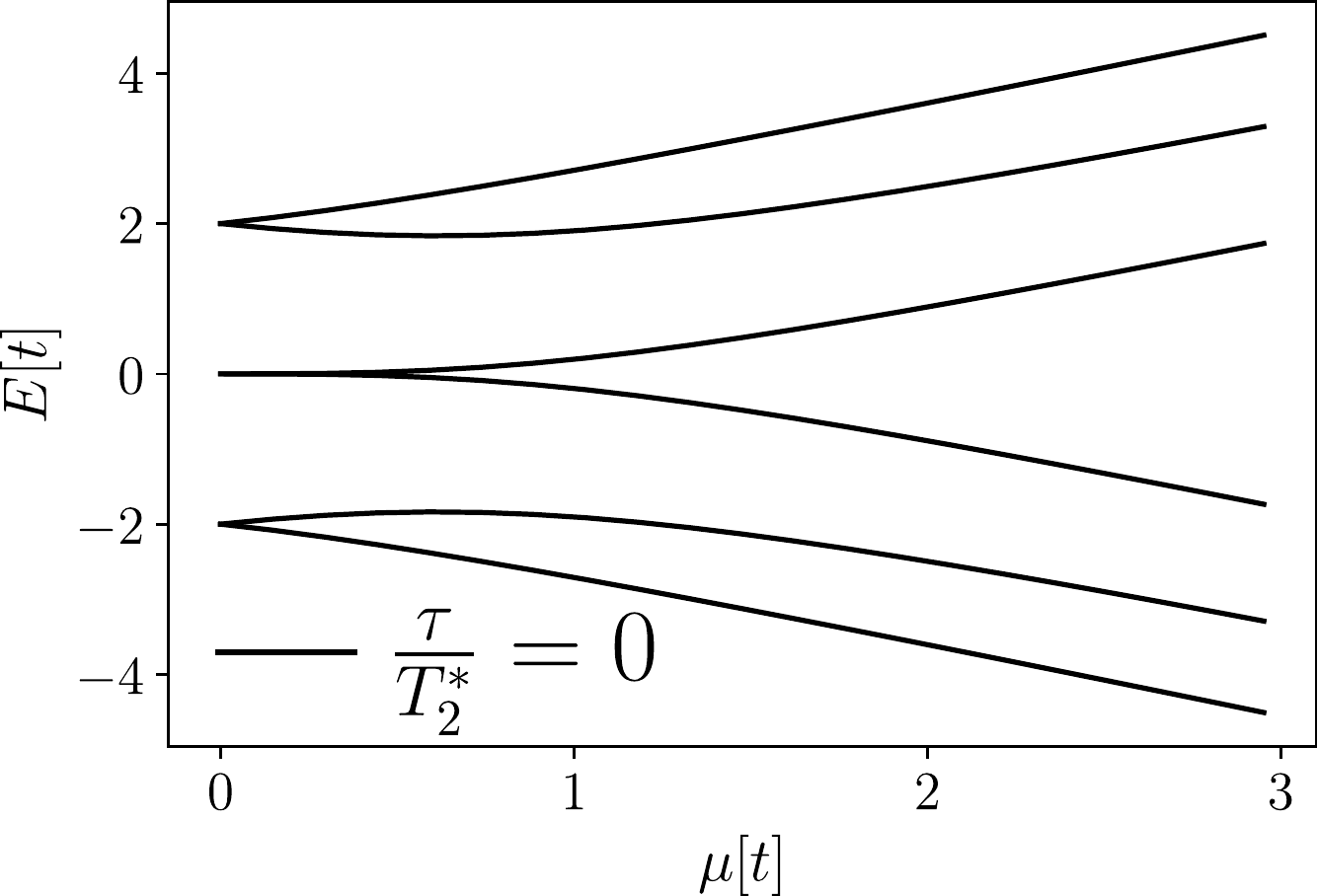}
		\end{minipage}\begin{minipage}{0.25\textwidth}
			\centering
			\includegraphics[width=1.0\textwidth]{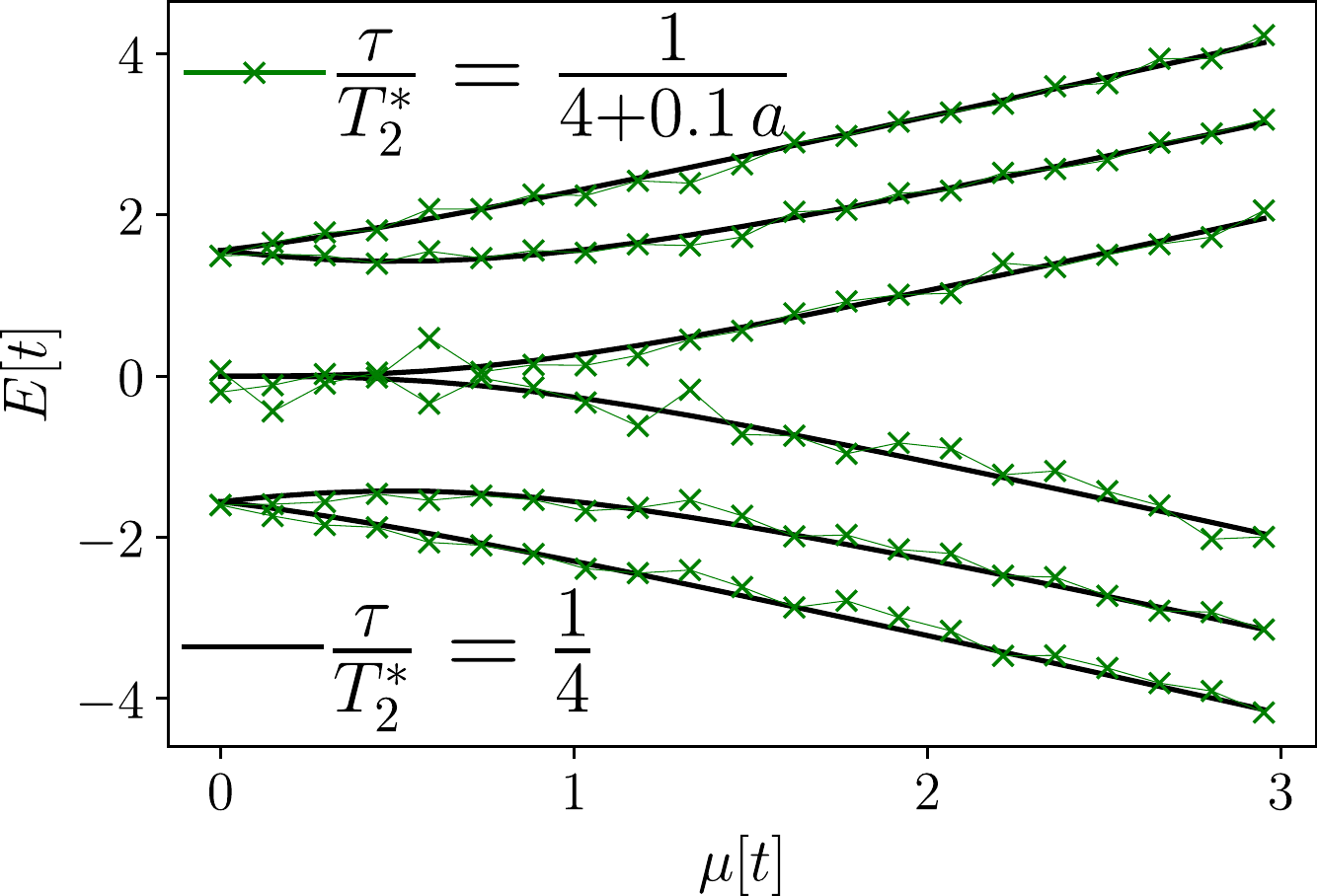}
		\end{minipage}\begin{minipage}{0.25\textwidth}
			\centering
			\includegraphics[width=1.0\textwidth]{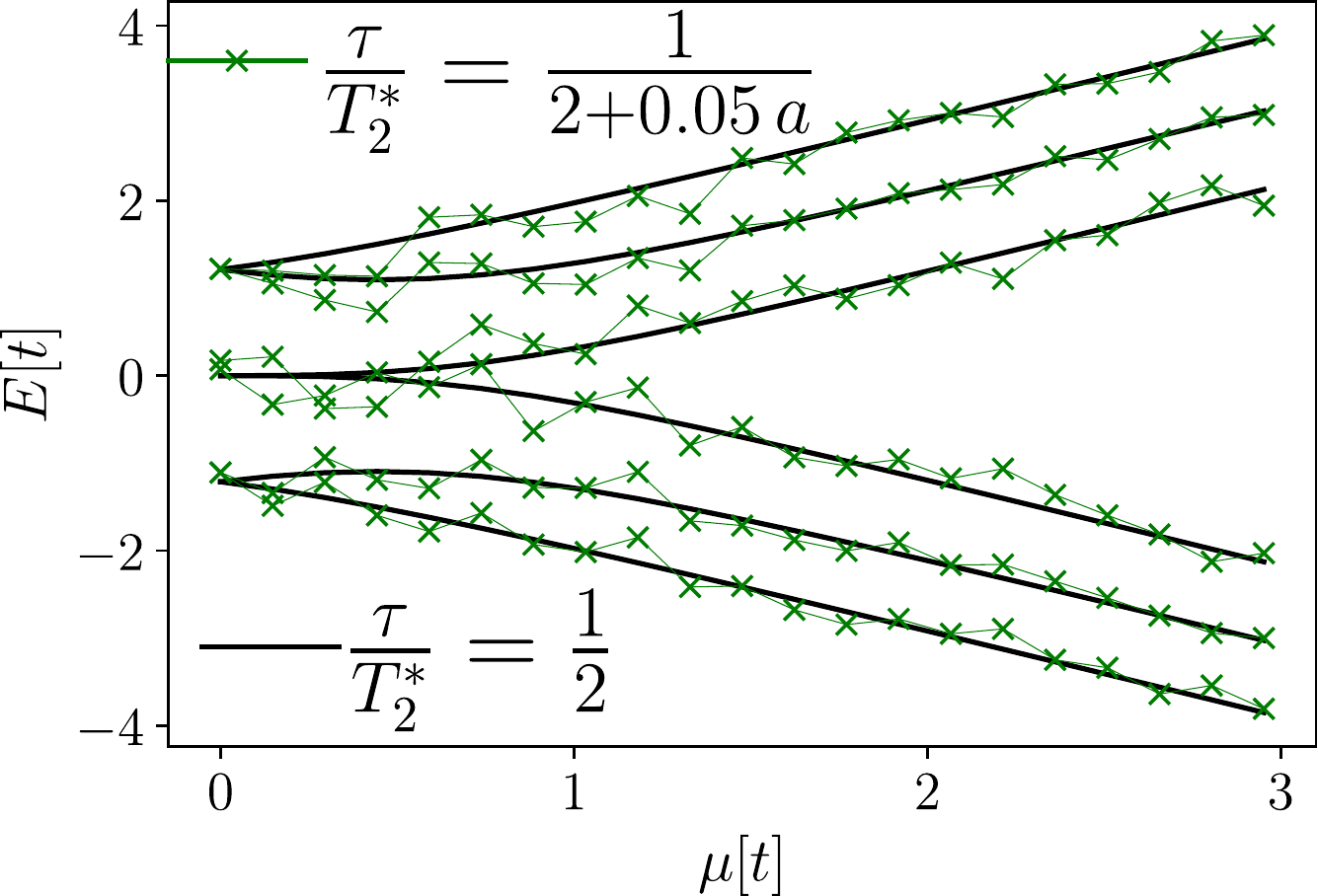}
		\end{minipage}\begin{minipage}{0.25\textwidth}
			\centering
			\includegraphics[width=1.25\textwidth]{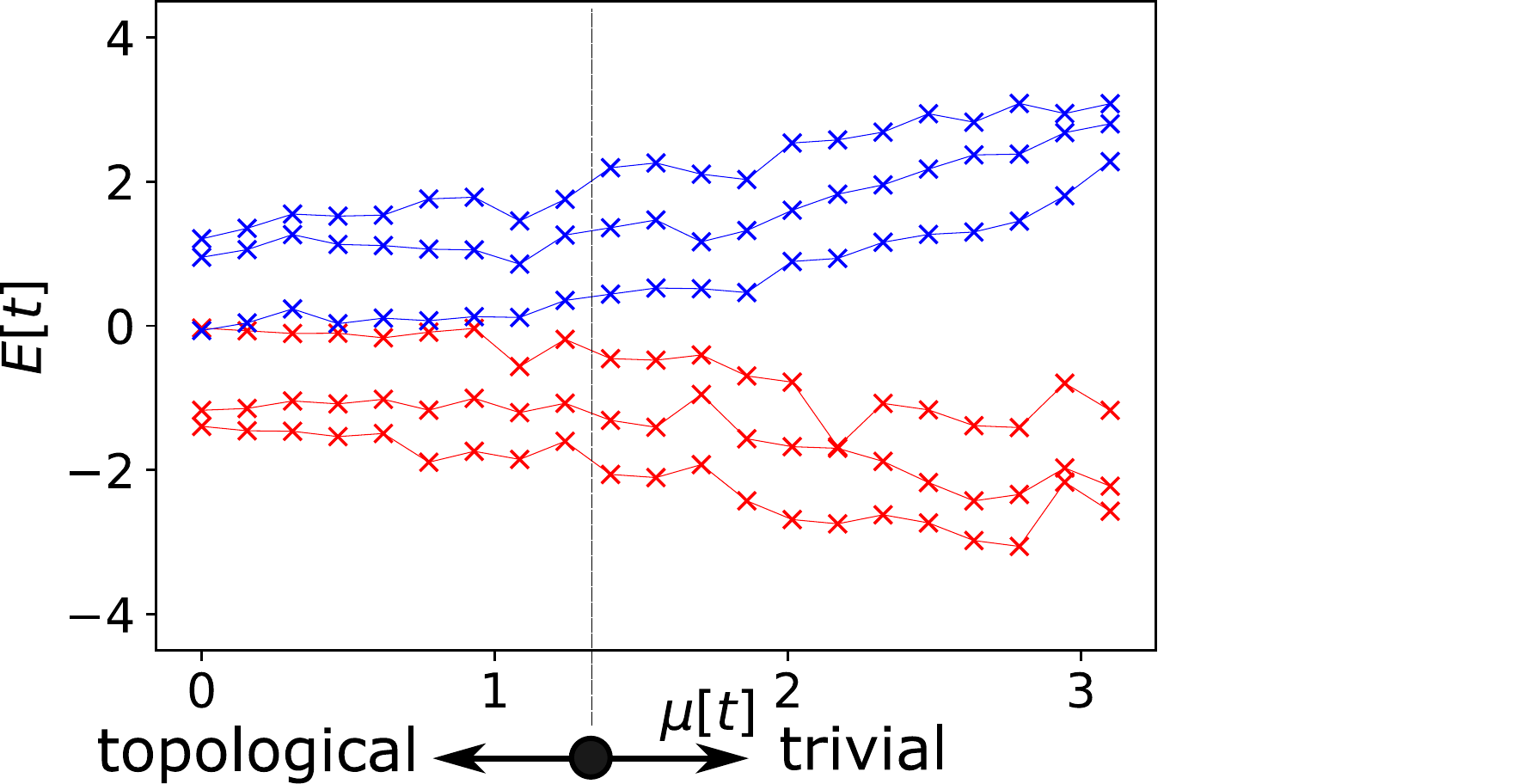}
		\end{minipage}
		\caption{First 3 figures - result of modeling, last figure  - experimental data. $a$ is a random number $a=[-1,1]$\label{fig:NumMod}}
	\end{figure*}
	
	Noise in a quantum system governed by the Hamiltonian $H$ and described by a density matrix $\rho$ is given by \cite{PhysRevB.93.205433}
	
	\begin{equation}\label{eq:Lindblad}
	{\dot{\rho}}=-\frac{i}{\hbar}[H,\rho]+\frac{1}{2}\sum_{ij}\Gamma_{ij}(2 L_{ij}^{\dagger}\rho L_{ij}-L_{ij} L_{ij}^{\dagger}\rho-\rho L_{ij} L_{ij}^{\dagger}).
	\end{equation}
	Here, $\Gamma_{ij}$ is the dissipation rate between the channel $ij$, and $ L_{ij}^{\dagger}=|i\rangle\langle j|$ and $L_{ij}=|j\rangle\langle i|$ are Lindblad dissipation operators. In general if $i\neq j$ the terms in the Lindblad equation describe relaxation processes happening on a characteristic relaxation timescale $T_1$ and if $i=j$ these terms correspond to pure dephasing processes described by a characteristic time $T_2^*$. In general $T_2^*\ll T_1$ -  pure dephasing dominates over relaxation - and this is the assumption I adopt in the remainder of the Section.
	
	The solution of a Lindblad equation in the presence of only pure dephasing is
	\begin{equation}
	\tilde{\rho}(\tau)=\rm{diag}\left(\rho(0)\right)+\sum_{i\neq j}\rho_{ij}e^{-\frac{\tau}{T_2^*}},
	\end{equation}
	the diagonal part off the density matrix remains unchanged while the off-diagonal terms are decaying with a factor of $\exp{\left(-\tau / T_2^*\right)}$ (see Section ''Phase damping" in Ref. \cite{nielsen2002quantum}). 
	
	By repeating the procedure from the main body of the paper - obtaining the quantum circuit with OpenFermion and then getting the ground state wavefunction $|\psi\rangle$ in a form of a numerical vector we can obtain the density matrix $\rho(0)=|\psi\rangle \langle \psi|$. After obtaining the density matrix and multiplying its off-diagonal terms by $\exp{\left(-\tau / T_2^*\right)}$ we obtain the noisy density matrix of the system $\tilde{\rho}$ (this simulates idle noise processes). The noisy expectation value of the Hamiltonian can be obtained as 
	
	\begin{equation}
	E=\rm{Tr}(\tilde{\rho} , H).
	\end{equation}
	
	Recent experimental and theoretical studies \cite{PhysRevLett.121.090502,PhysRevLett.123.190502,burnett2019decoherence, etxezarreta2021time} indicate that the dephasing time of superconducting qubits varies in time. A minimal model in describing that is assigning random fluctuations of the dephasing time between measurements of different eigenstates of energy at different values of $\mu$, $T_2^*\rightarrow T_2^*+ab$ where $a=[-1,1]$ is a random number and $b\ll T_2^*$.
	
	In Fig. \ref{fig:NumMod} I show what is the effect of pure dephasing on the spectrum of MZMs. As the duration of the quantum circuit is increased, the circuit dephases more (first to third subfigure) the topologically trivial states ($E=\pm 2$ at $\mu=0^+$) move more towards zero in the BdG spectrum (black lines). Furthermore, variations in the dephasing time between the realization of the experiment yield a qualitatively similar spectrum to the one measured (green symbols in the 3rd subfigure vs. measured data - 4th subfigure). This effect is visually similar to a slight violation of particle-hole symmetry.
	
	\section{S4 - The Majorana edge correlation function}
	
	In the thermodynamic limit, the Majorana edge correlation function is expected to decay with a quadratic dependence in $\mu$ \cite{miao2018majorana}. As comparing a $n=3$ site Kitaev chain with a theory developed for an infinitely long chain is infeasible, I will just give mean square fits of the decay of the absolute value of the Majorana edge correlation function $|\langle i \gamma_1 \gamma_k \rangle |$, with a goal of outlining the fact that Majorana edge correlations do indeed decay as a function of $\mu$. The polynomial fit to data for the first run (left) for state $[0]$ is $0.57-0.18\mu+0.03\mu^2$ and for the state $[1,2]$ is $0.56-0.18\mu +0.03\mu^2 $ with respective residuals $0.03$ and $0.02$. 
	
	The polynomial fit to data for the second run (right, main body of the paper) for state $[0]$ is $0.63-0.14\mu+0.03 \mu^2$ and for state $[1,2]$ $0.6-0.09\mu+0.01\mu^2$ with residuals of $0.01$ and $0.03$ respectively. The device was calibrated between runs. 
	
	\section{S5 - Comparison between runs of hardware}
	
\begin{figure*}
		\centering
		\begin{minipage}{0.25\textwidth}
			\centering
			\includegraphics[width=1.0\textwidth]{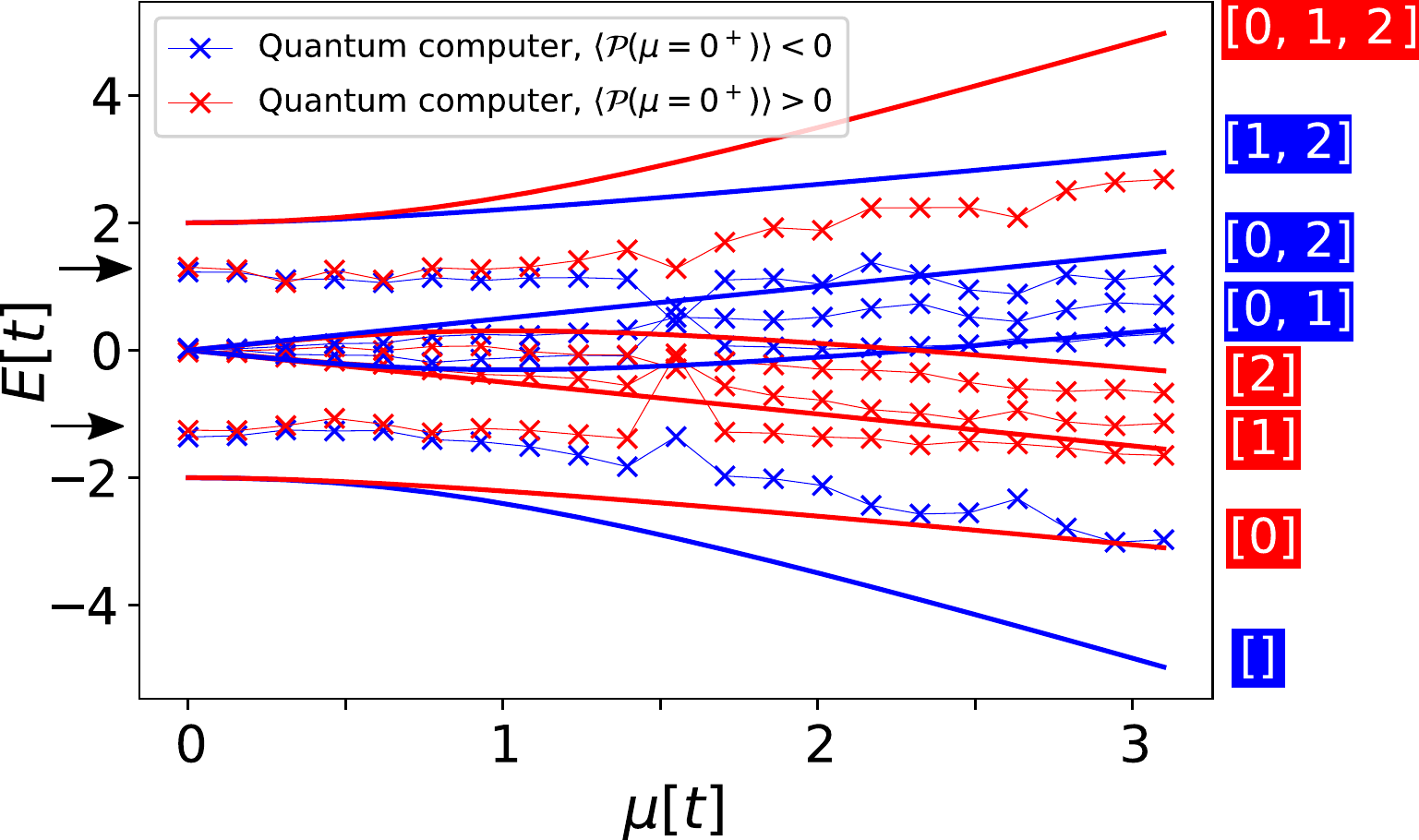}
		\end{minipage}\begin{minipage}{0.25\textwidth}
			\centering
			\includegraphics[width=1.0\textwidth]{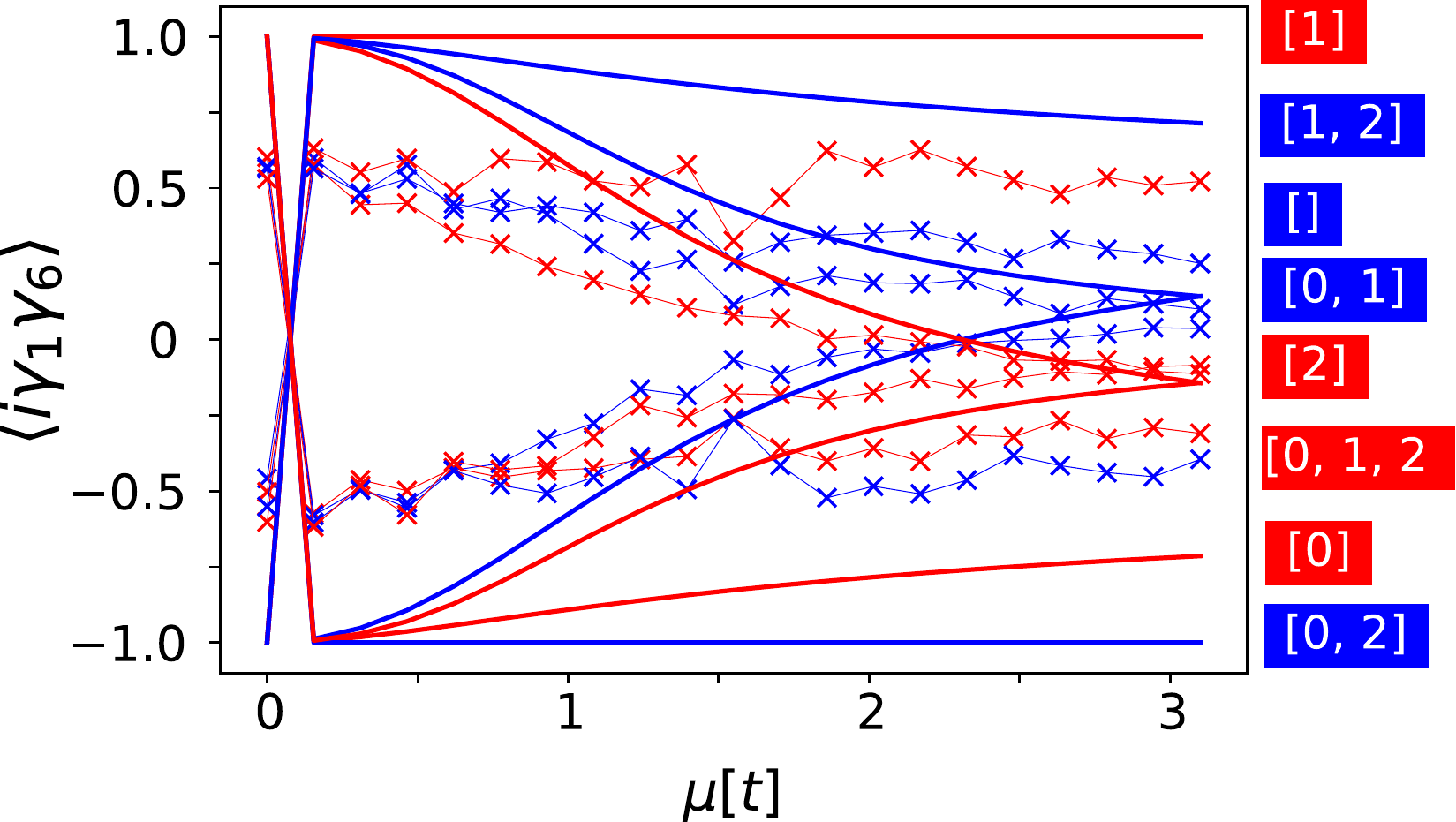}
		\end{minipage}\begin{minipage}{0.25\textwidth}
			\centering
			\includegraphics[width=1.0\textwidth]{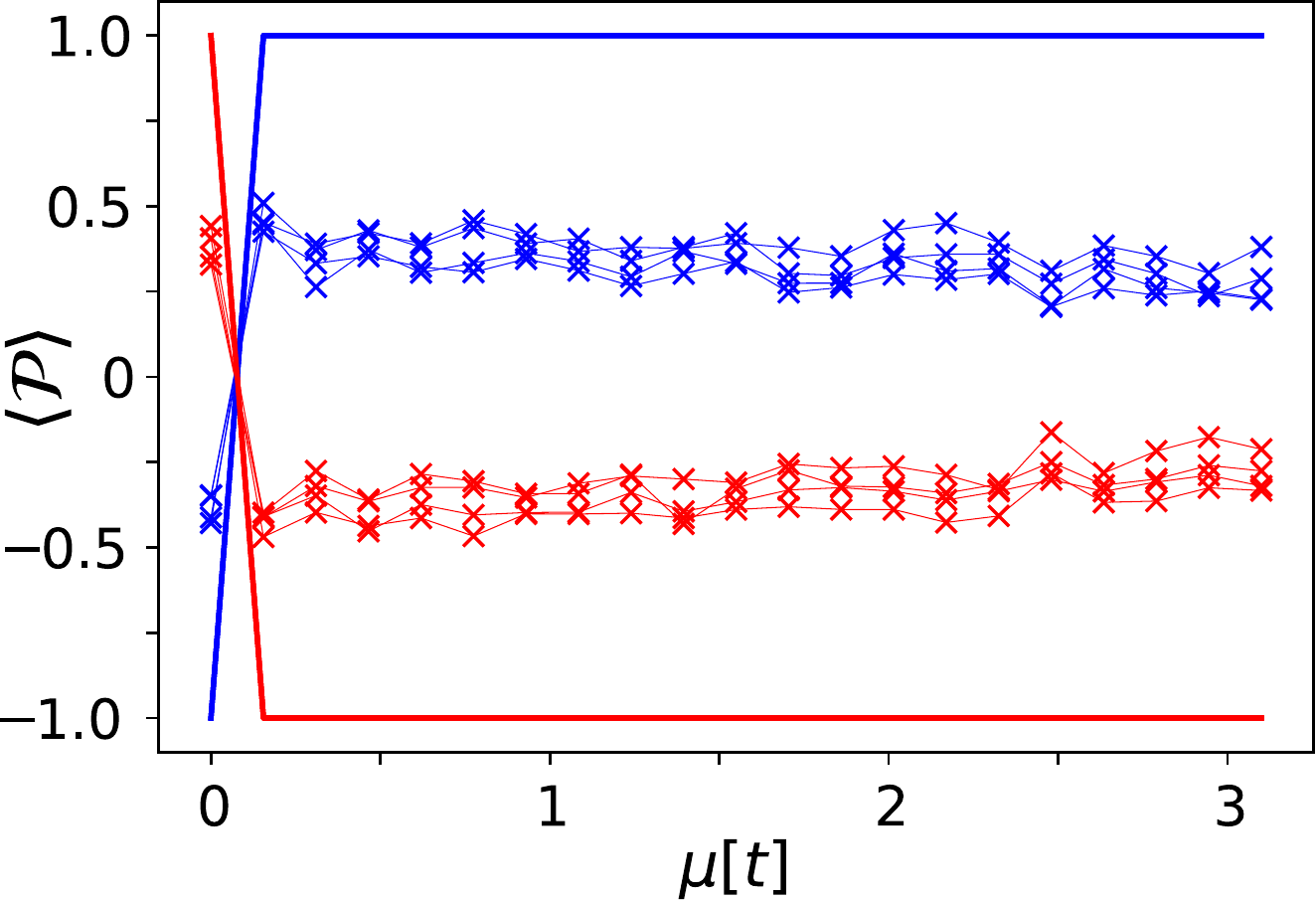}
		\end{minipage}\begin{minipage}{0.25\textwidth}
			\centering
			\includegraphics[width=1.0\textwidth]{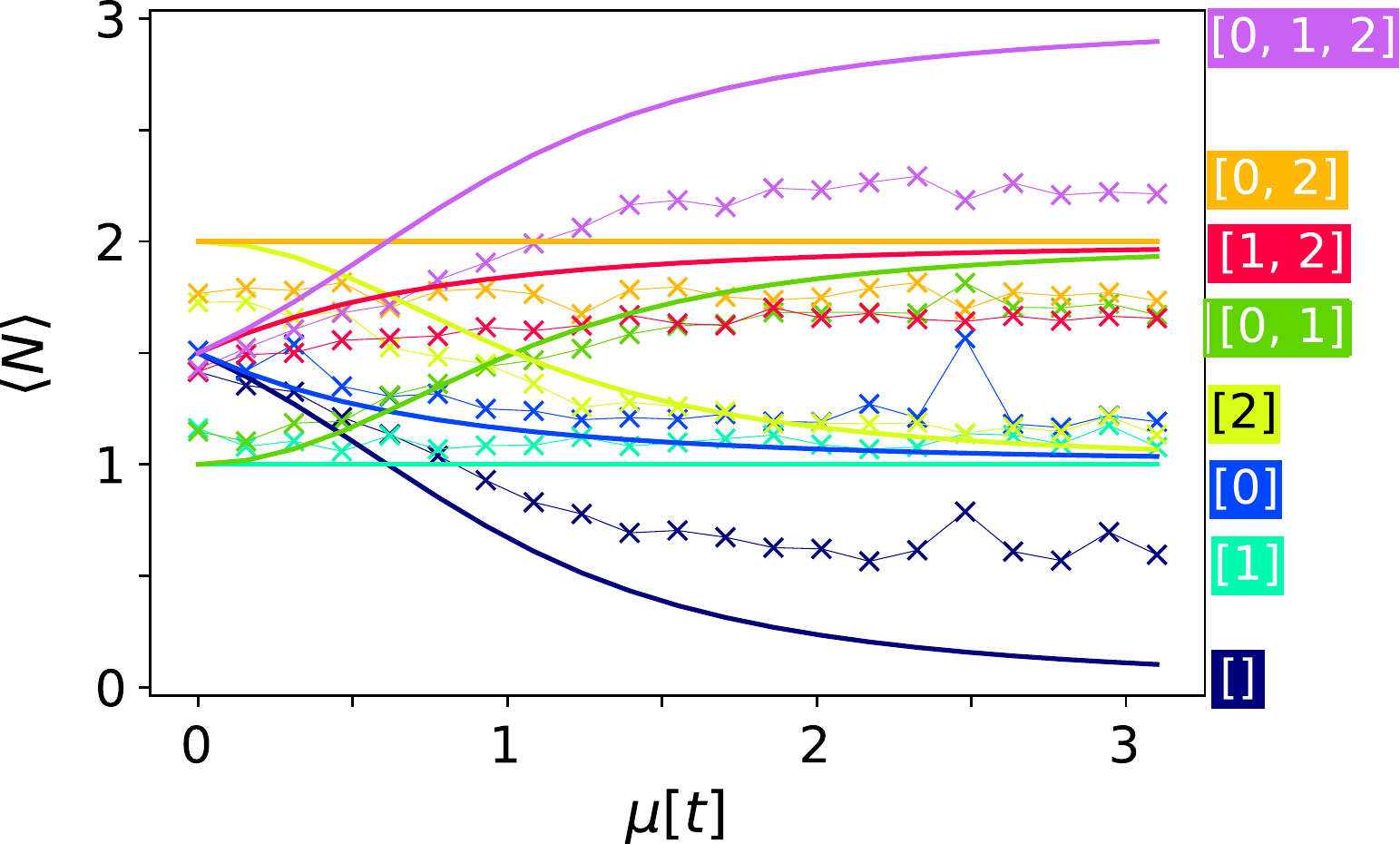}
		\end{minipage}
		\begin{minipage}{0.25\textwidth}
			\centering
			\includegraphics[width=1.0\textwidth]{Gauss_vs_QC_3.pdf}
		\end{minipage}\begin{minipage}{0.25\textwidth}
			\centering
			\includegraphics[width=1.0\textwidth]{corr_3.pdf}
		\end{minipage}\begin{minipage}{0.25\textwidth}
			\centering
			\includegraphics[width=1.0\textwidth]{par_3.pdf}
		\end{minipage}\begin{minipage}{0.25\textwidth}
			\centering
			\includegraphics[width=1.0\textwidth]{num_3.pdf}
		\end{minipage}
		\caption{TOP - first realisation of: Energy, Majorana edge correlation function, parity and particle number respectively as a function of chemical potential. BOTTOM - second realisation (main body of paper) of: Energy, Majorana edge correlation function, parity and particle number respectively as a function of chemical potential. Full lines denote ideal simulation, x-points connected with lines are a result of an experiment on IBM Santiago.\label{fig:comp}}
	\end{figure*}

\begin{figure*}
		\begin{minipage}{0.5\textwidth}
			\includegraphics[width=0.9\textwidth]{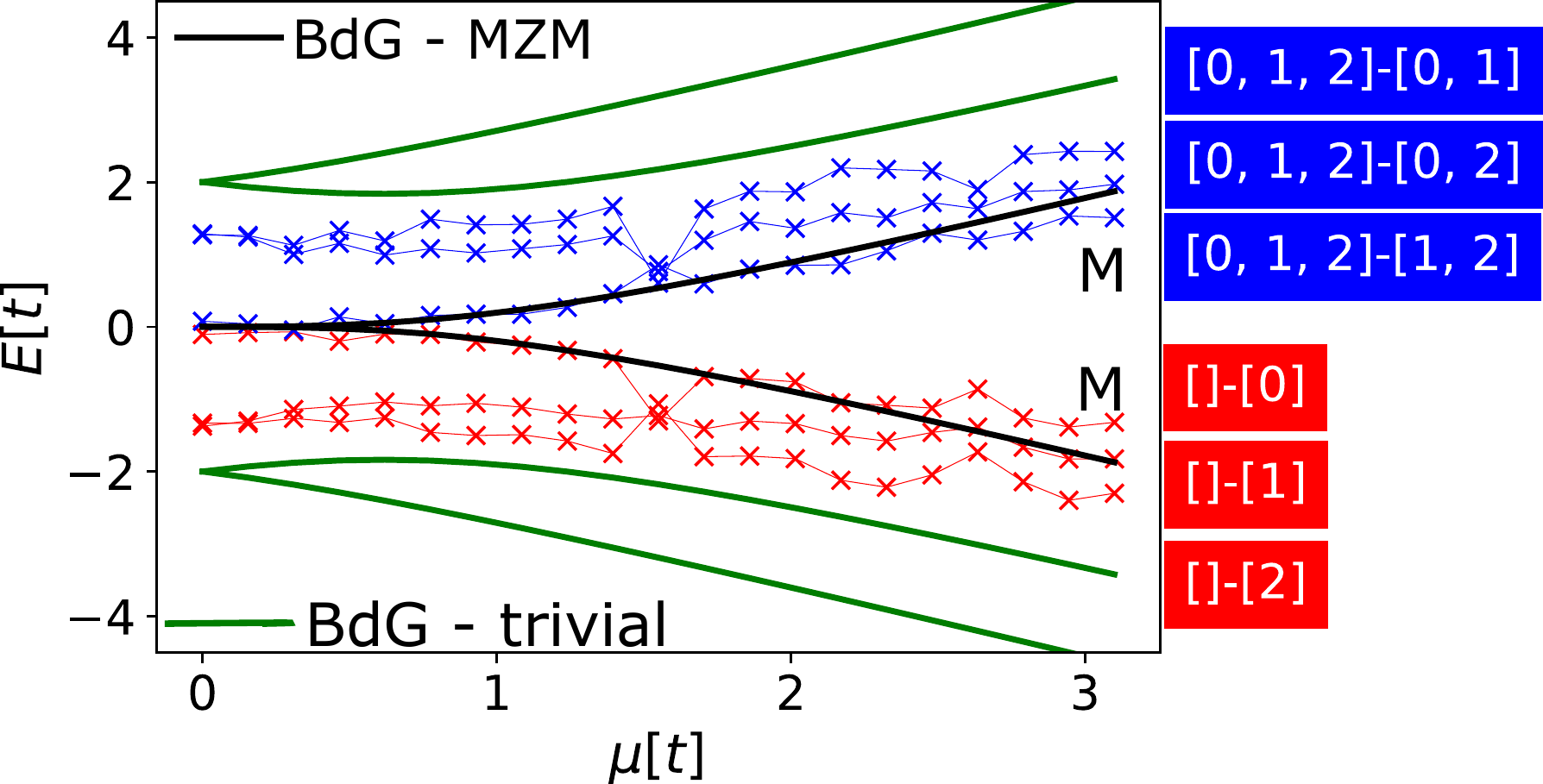}
			
		\end{minipage}\begin{minipage}{0.5\textwidth}
			\includegraphics[width=0.9\textwidth]{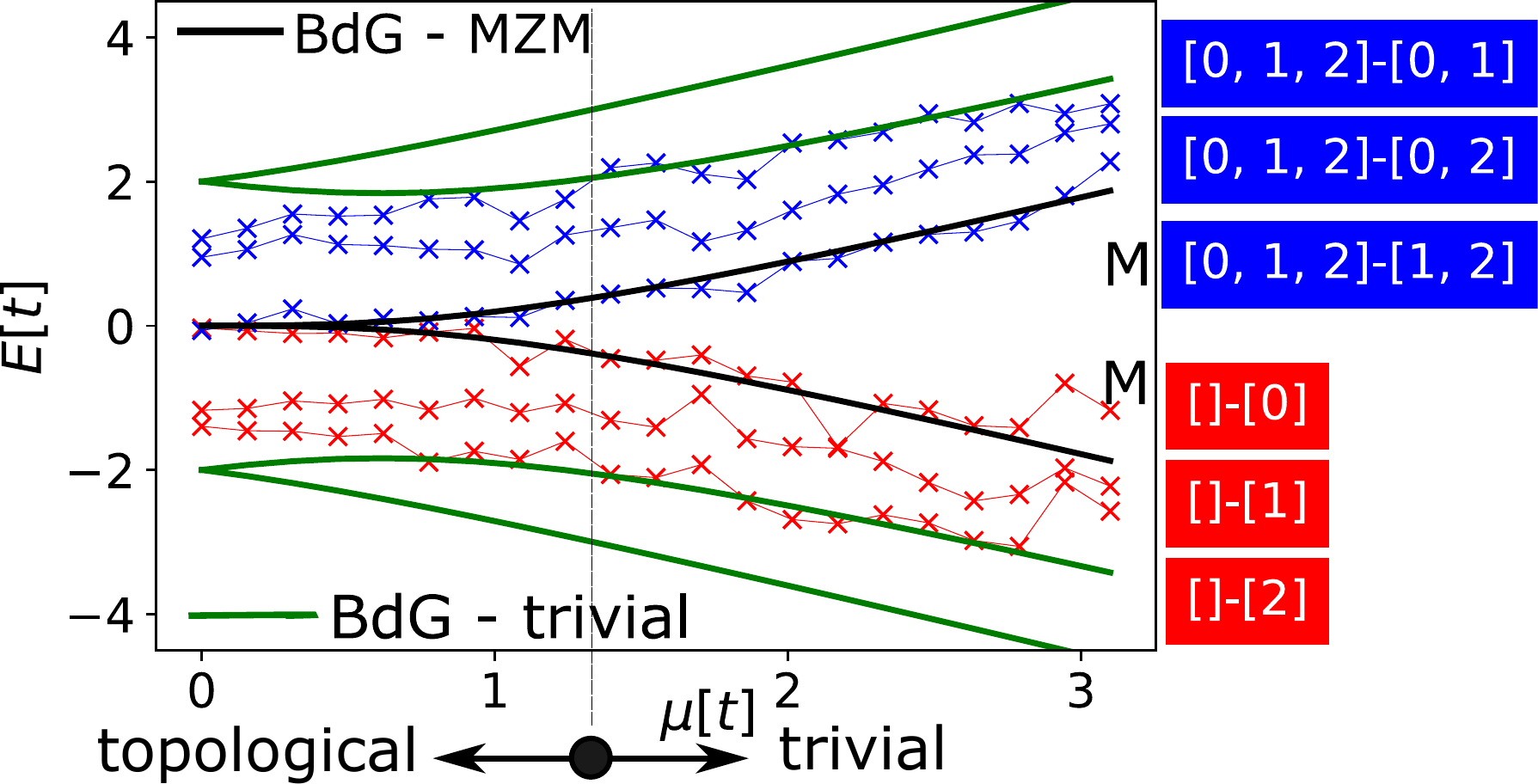}
		\end{minipage}
		\caption{The BdG Hamiltonian theory (black and green) and experiment (blue and red). LEFT first realisation of the experiment, RIGHT second realisation of the experiment (main body of the paper). \label{fig:bdg_comp}}
	\end{figure*}

	In this section I compare two different experiment runs on IBM Santiago in FIG. \ref{fig:comp}. Although the different executions vary slightly quantitatively depending on the calibration of the device all qualitative features remain present, especially the possible topological degeneracy. In Fig. \ref{fig:bdg_comp} I display a comparison between a BdG spectrum and two different realizations of the experiment. Both experiments show excellent qualitative agreement with the BdG prediction. One can quantify this by defining a mean absolute error as ${\rm ME}=\left(\sum_i |x_i-y_i|\right)/m$, where $m$ is the total number of measurements/predictions, $x_i$ is the $i$th measurement of energy from the quantum device and $y_i$ is the $i$th prediction of the BdG Hamiltonian. I find a ${\rm ME}=0.127$ for the left figure and ${\rm ME}=0.129$ for the right figure for $-1.87 \le y_i\le 1.87$.
	
	\section{S6 - A $4$-site Kitaev chain}
	
	The formalism presented here allows the treatment of longer Kitaev chains with quantum simulators.  Actual quantum computing preparation of Kitaev states should somewhat still be possible with currently available quantum computing hardware, however the author of this paper has no access to state of the art quantum computers and I will thus only focus on results from quantum simulators in the remainder of this Supplementary Material.
	
	In FIG. \ref{fig:overview4} I compare results obtained by an ideal simulation of Gaussian states with BdG energies (a) and display the eigenspectrum obtained on a quantum computing simulator (b).
	
	In subfigure (c) Majorana edge-correlation functions are shown with states $[0]$ and $[1,2,3]$ having a decay of the edge correlation function. Parity switches are also in accordance with single-particle picture theories. In FIG. \ref{fig:paritysn4} I show a matching between parity switches predicted by Gaussian states implemented on an ideal quantum computing simulator and white lines are predictions of Eq. (2). Grey areas are ranges in parameter space of $\mu$ where parity switching exists due to a finite resolution of $\mu$.
	
	Also in similarity with the 3-site case Majorana zero modes start favoring pairing between neighboring Majoranas as the chemical potential is increased see states $[0,2,3]$ and $[1]$ in FIG. \ref{fig:corr4}. 

	\begin{figure*}[b!]
		\begin{minipage}{0.5\textwidth}
			\includegraphics[width=0.9\textwidth]{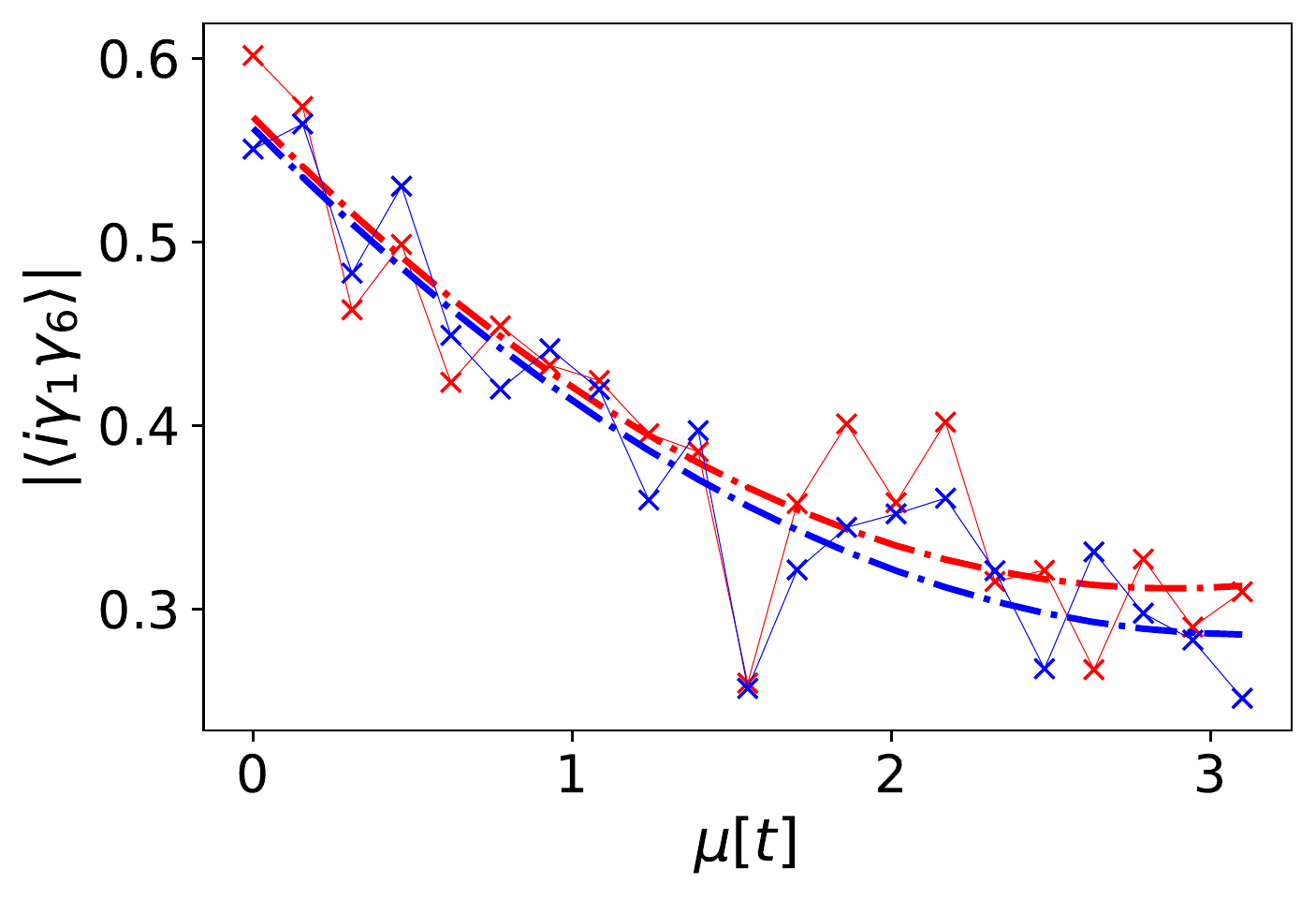}
		\end{minipage}\begin{minipage}{0.5\textwidth}
			\includegraphics[width=0.9\textwidth]{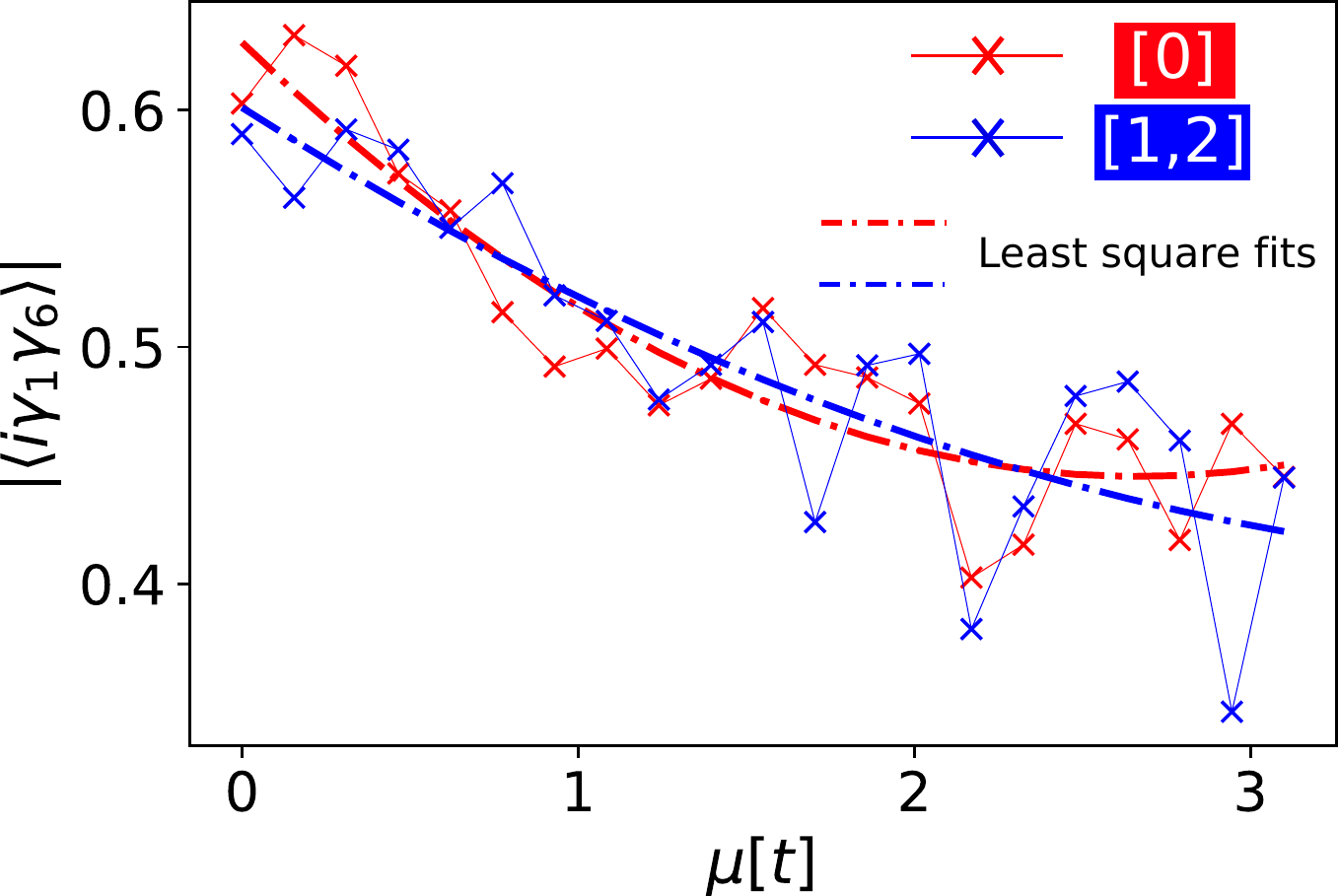}
		\end{minipage}
		\caption{The absolute value of the correlation function of Majorana zero modes. LEFT first realization of the experiment, RIGHT second realization of the experiment (main body of the paper). \label{fig:cor_zoom}}
	\end{figure*}
	
	\begin{figure*}[b!]
		\begin{minipage}{0.48\textwidth}
			\centering  
			\subfloat{(a)}{		
				
				\includegraphics[height=4.5cm]{BdG_4.pdf}}
		\end{minipage}\begin{minipage}{0.48\textwidth}
			\centering
			\subfloat{(b)}{
				
				\includegraphics[height=4.2cm]{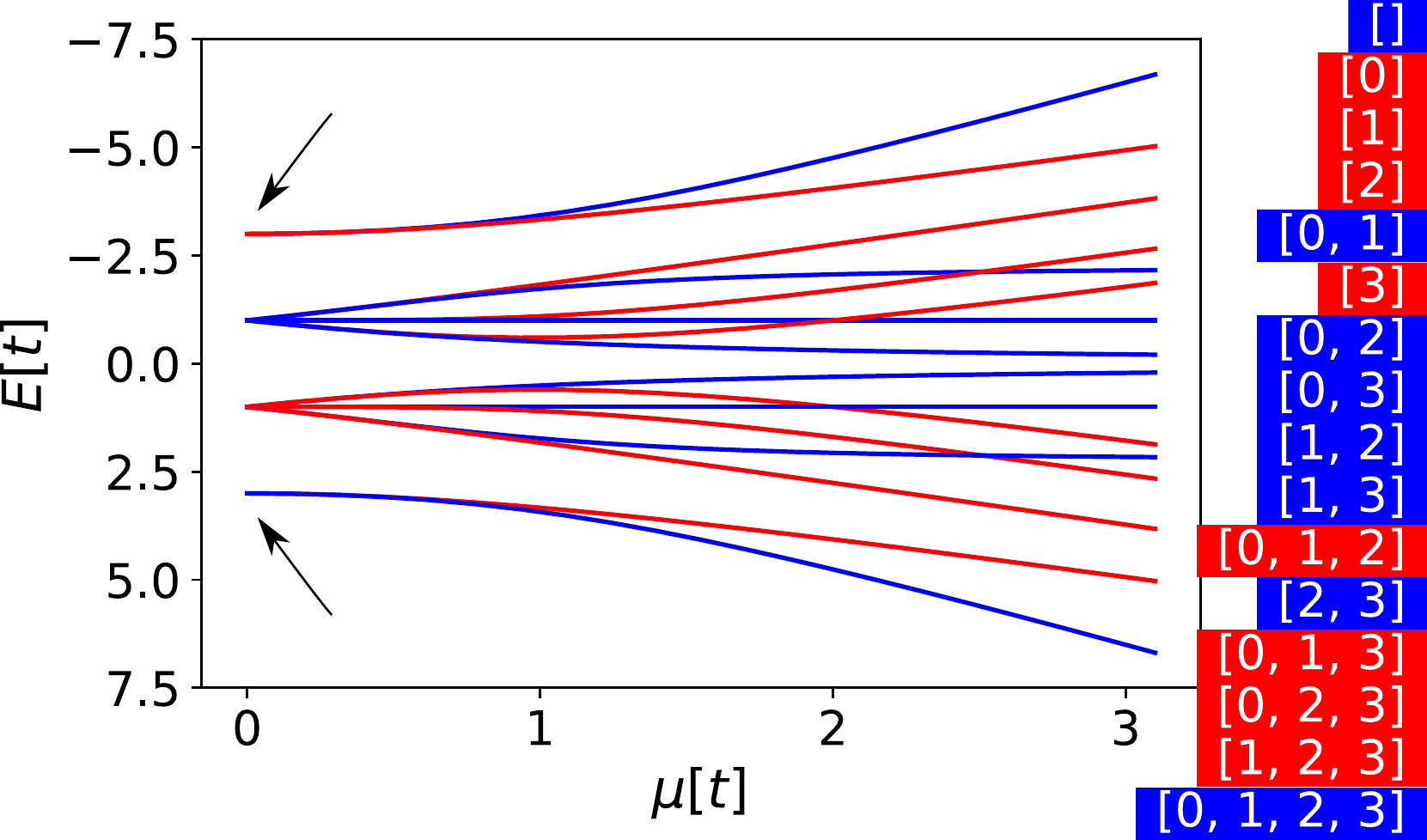}}
		\end{minipage}
		\begin{minipage}{0.48\textwidth}
			\centering
			\subfloat{(c)}{
				
				\includegraphics[height=4.2cm]{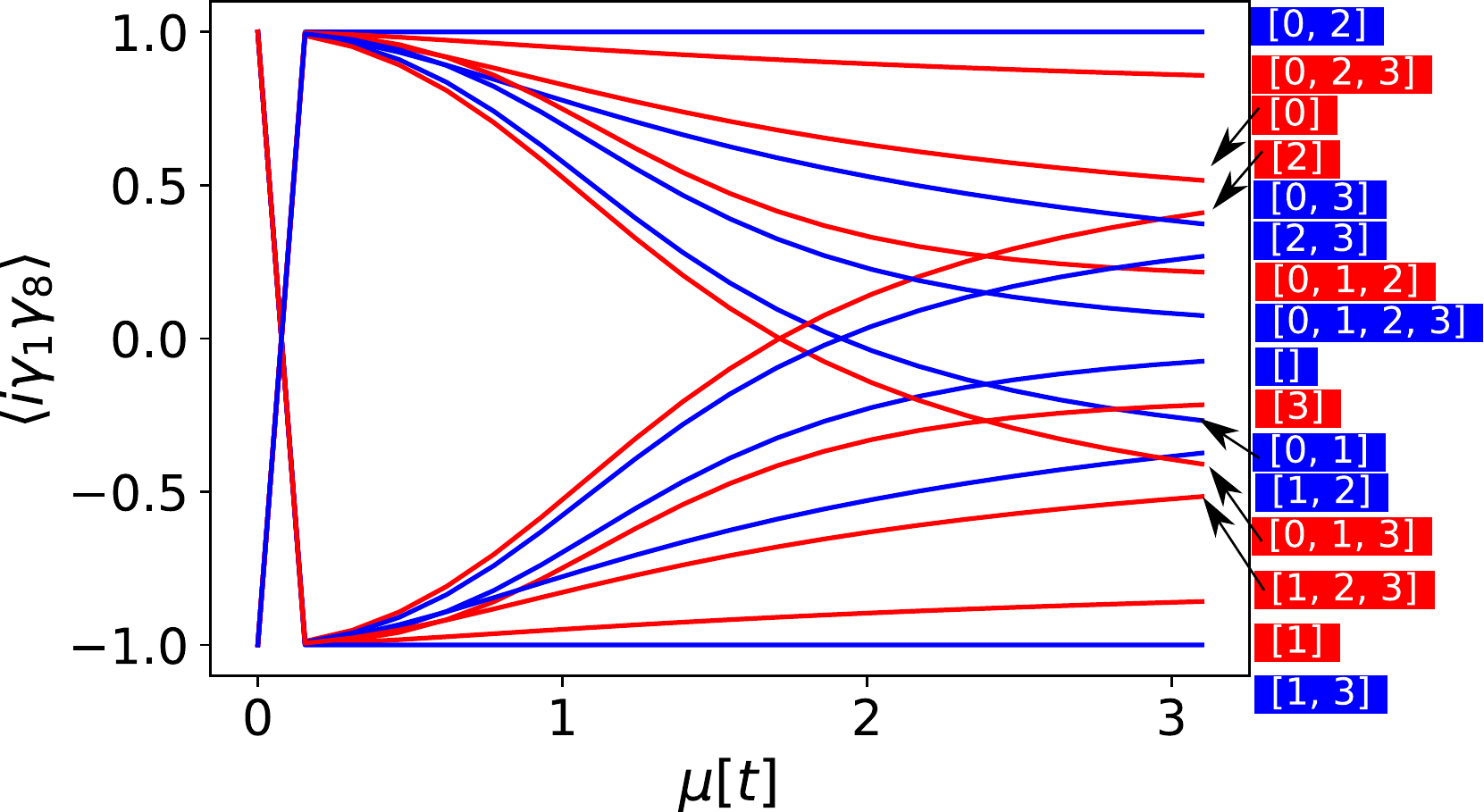}}
			
		\end{minipage}\begin{minipage}{0.48\textwidth}
			\centering
			
			\subfloat{(d)}{
				
				\includegraphics[height=4.2cm]{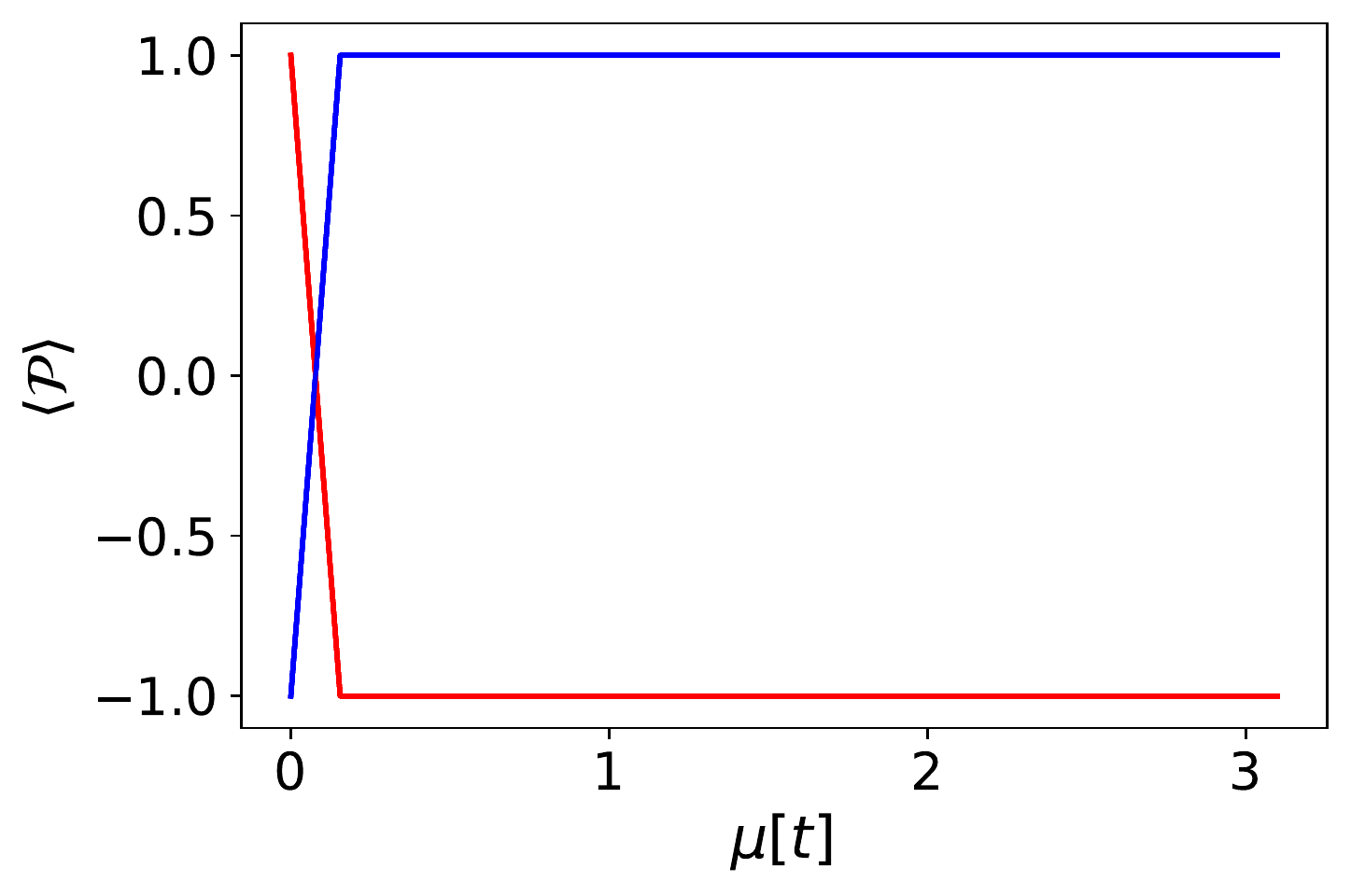}}
		\end{minipage}
		\caption{Quantities of a $4$-site Kitaev Hamiltonian at $t=-1$ and $\Delta=1$ as a function of the chemical potential $\mu$ in units of absolute value of tunnelling $[t]$. Red denotes values obtained on an ideal simulator of quantum computers with $\langle P(\mu=0^+)\rangle=1$. Blue denotes values obtained on an ideal simulator of quantum computers with $\langle P(\mu=0^+)\rangle=-1$. (a) The single-particle picture BdG spectrum, diagonalizatioin of the BdG Hamiltonian (grey lines) compared to Gaussian state solutions (coloured markers). (b) The full spectrum computed by Gaussian states. Black arrows indicate the topological degeneracy. (c) Majorana edge-correlation function, an ideal simulation with fermionic Gaussian states implemented on a quantum computing simulator. (d) Parity - an ideal simulation with Gaussian states. \label{fig:overview4}}
	\end{figure*}
	\begin{figure*}[b!]
		\begin{minipage}{0.33\textwidth}
			\subfloat{(a)}{	
				
				\includegraphics[height=3.9cm]{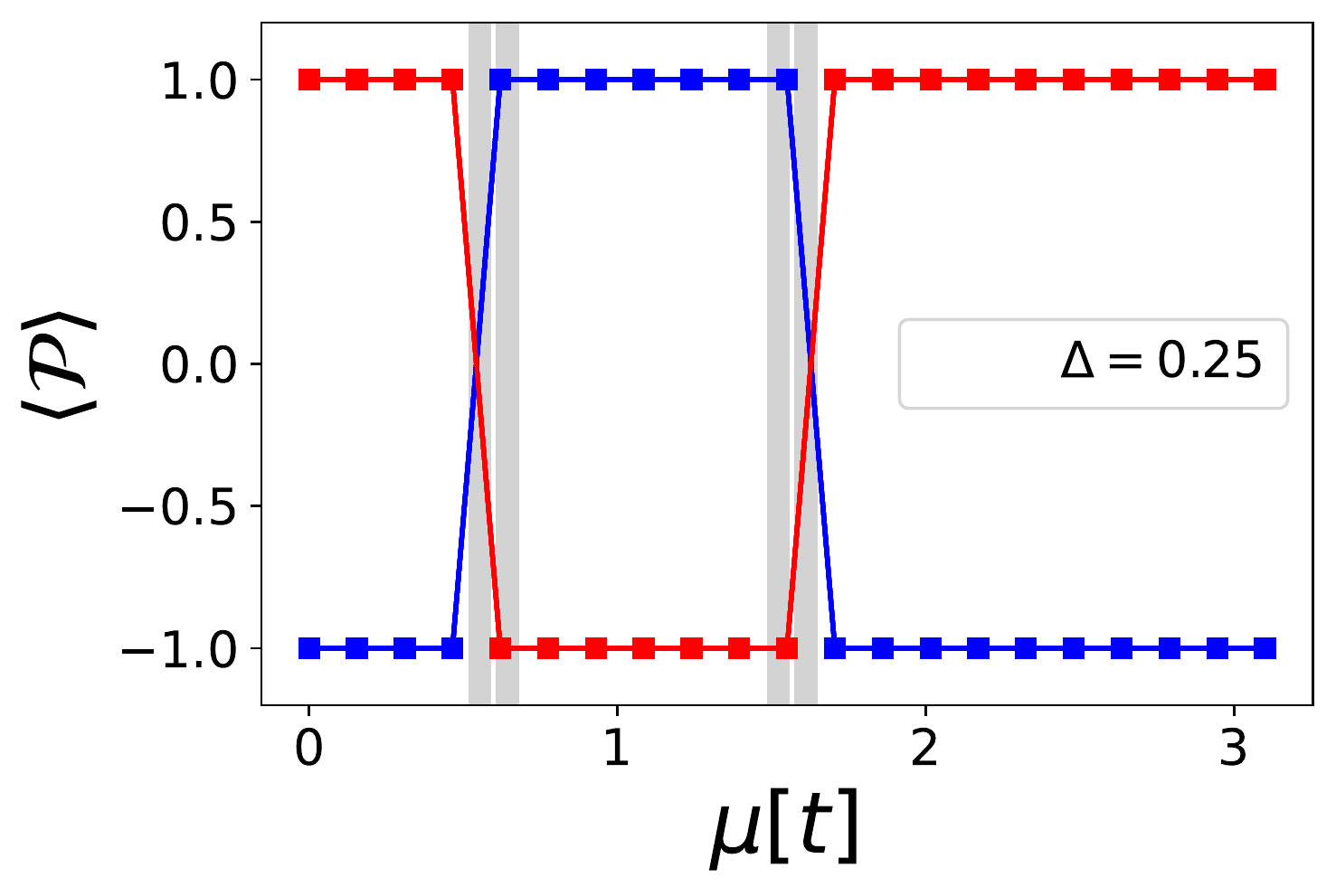}}
		\end{minipage}\begin{minipage}{0.33\textwidth}
			\subfloat{(b)}{	
				
				\includegraphics[height=3.9cm]{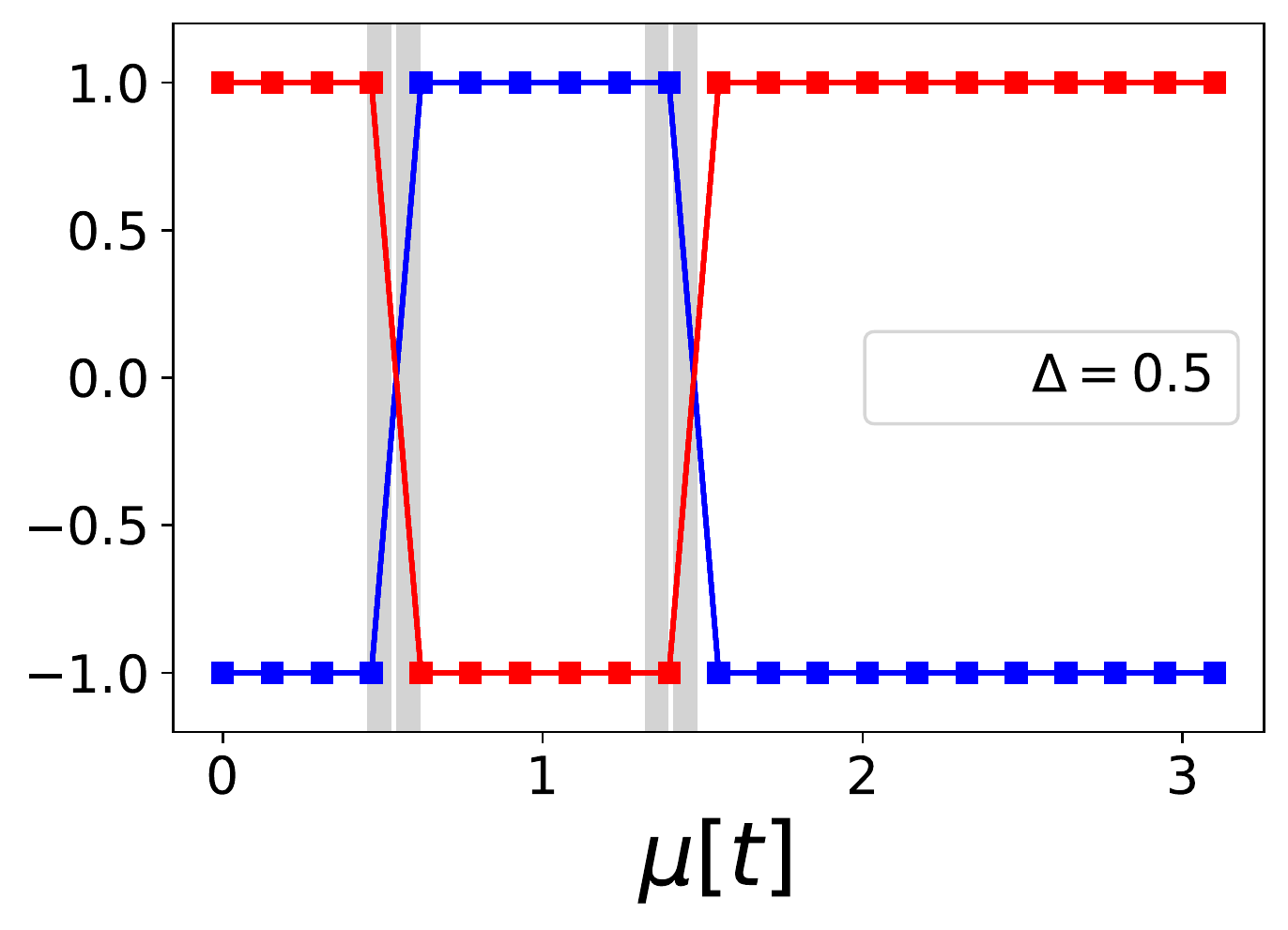}}
		\end{minipage}\begin{minipage}{0.33\textwidth}
			\subfloat{(c)}{	 
				
				\includegraphics[height=3.9cm]{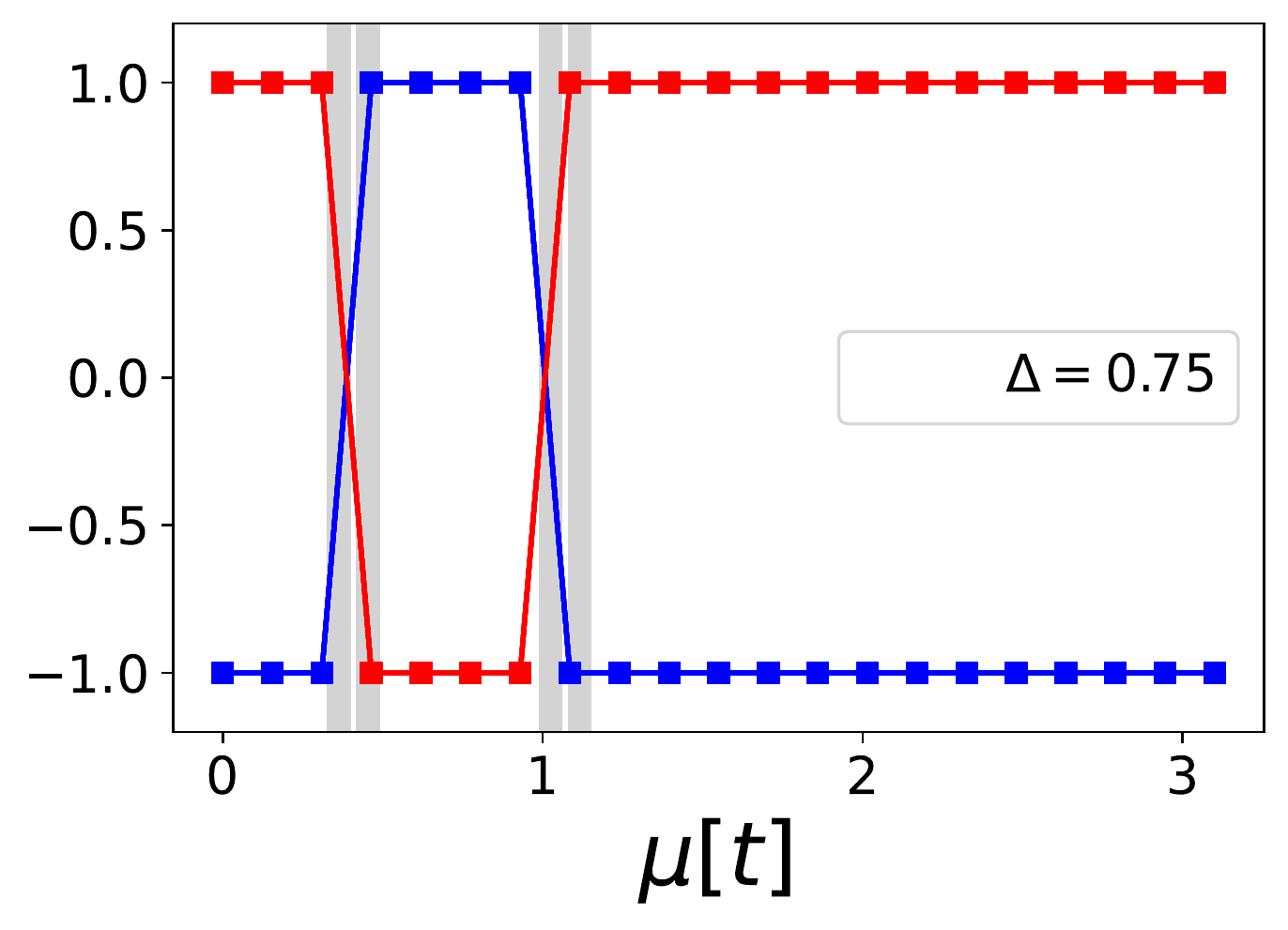}}
		\end{minipage}
		
		\begin{minipage}{0.33\textwidth}
			\subfloat{(d)}{	
				
				\includegraphics[height=3.9cm]{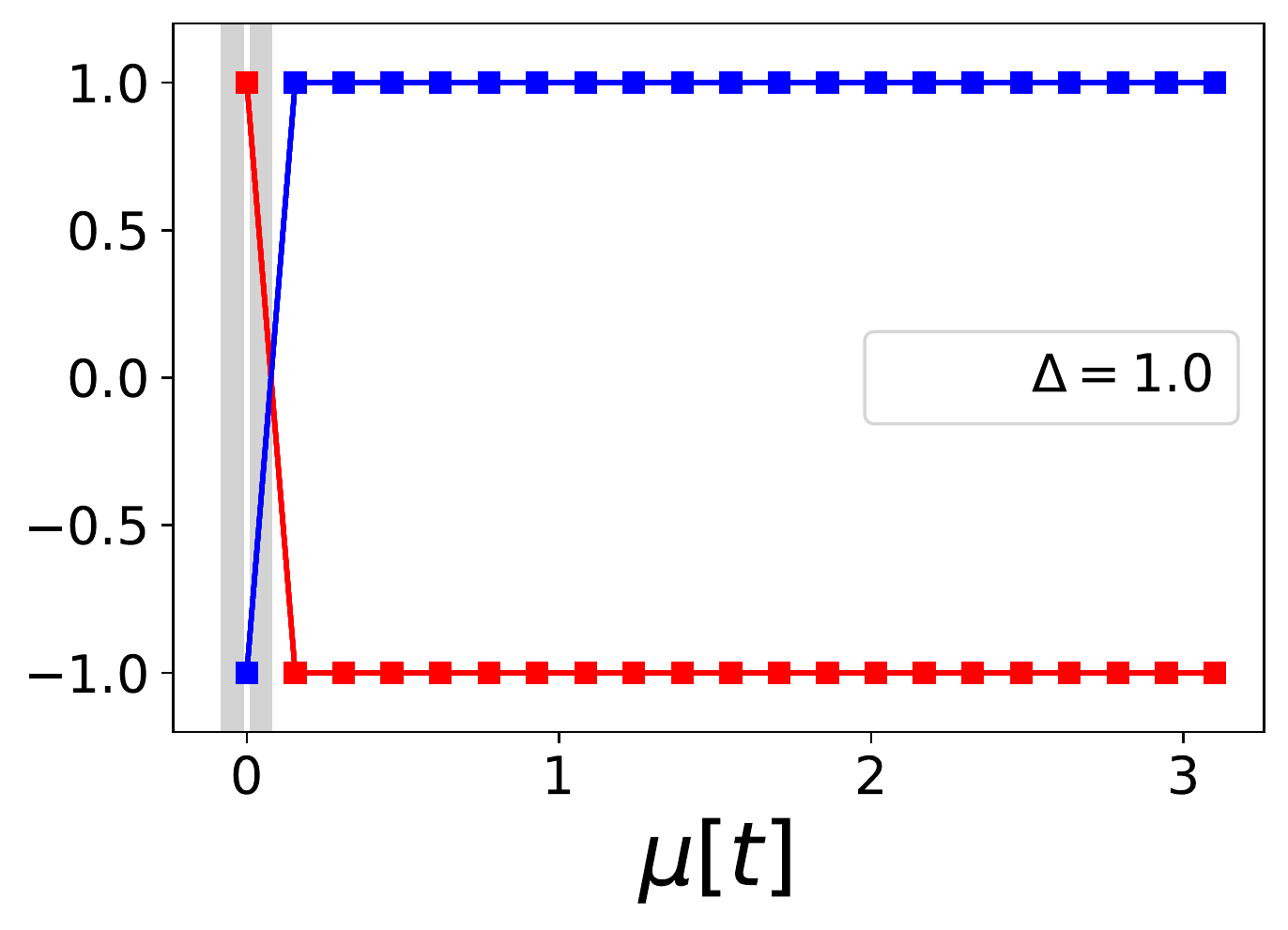}}
		\end{minipage}\begin{minipage}{0.33\textwidth}
			\subfloat{(e)}{	
				
				\includegraphics[height=3.9cm]{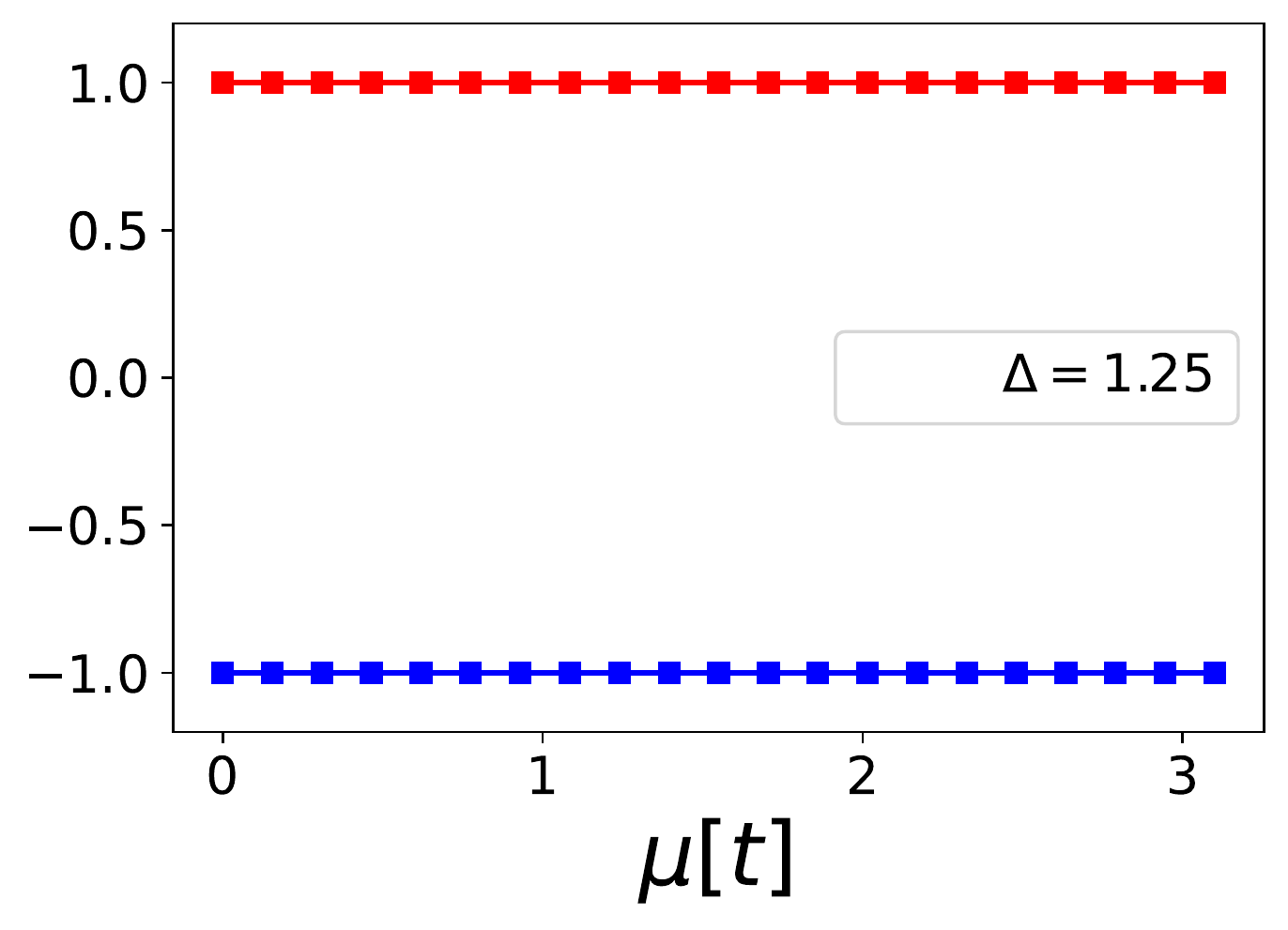}}	
		\end{minipage}	
		
		\caption{(a-e) Parity switches at $t=-1$ for different values of $\Delta$ and for 20 values of $\mu$ between $10^{-8}$ and $3.1$ with a $0.155$ increment for an $n=4$ Kitaev chain. White lines are predictions of Eq. (2) and grey areas are the regions in which parity switches potentially exist due to finite resolution in $\mu$. \label{fig:paritysn4}}. 
	\end{figure*}
	
	\begin{figure*}
		\begin{minipage}{0.33\textwidth}
			\centering  
			\subfloat{(a) $\mu=10^{-8}$}{	
				
				\includegraphics[height=3.7cm]{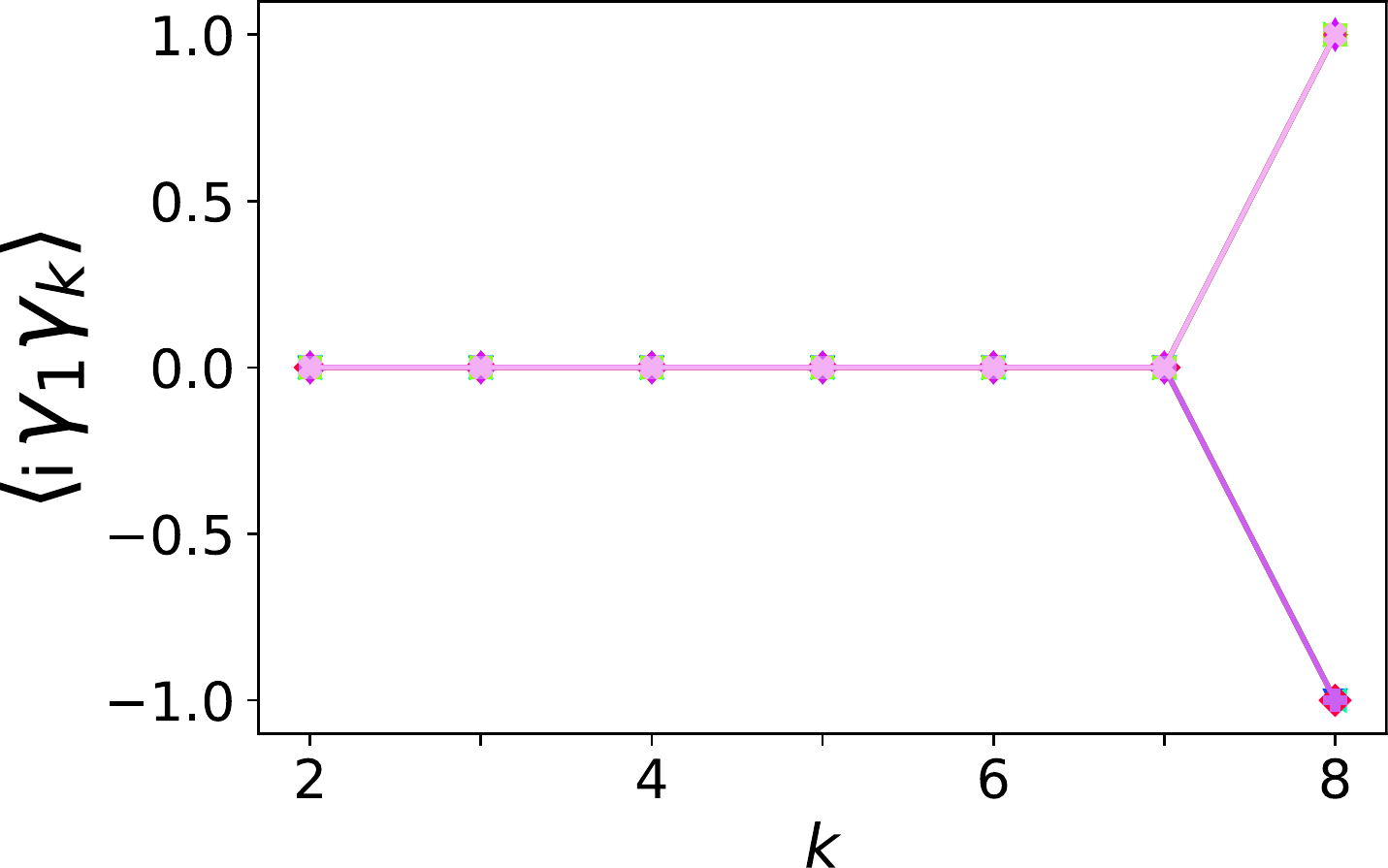}}
		\end{minipage}\begin{minipage}{0.33\textwidth}
			\subfloat{(b) $\mu=1$}{	
				
				\includegraphics[height=3.7cm]{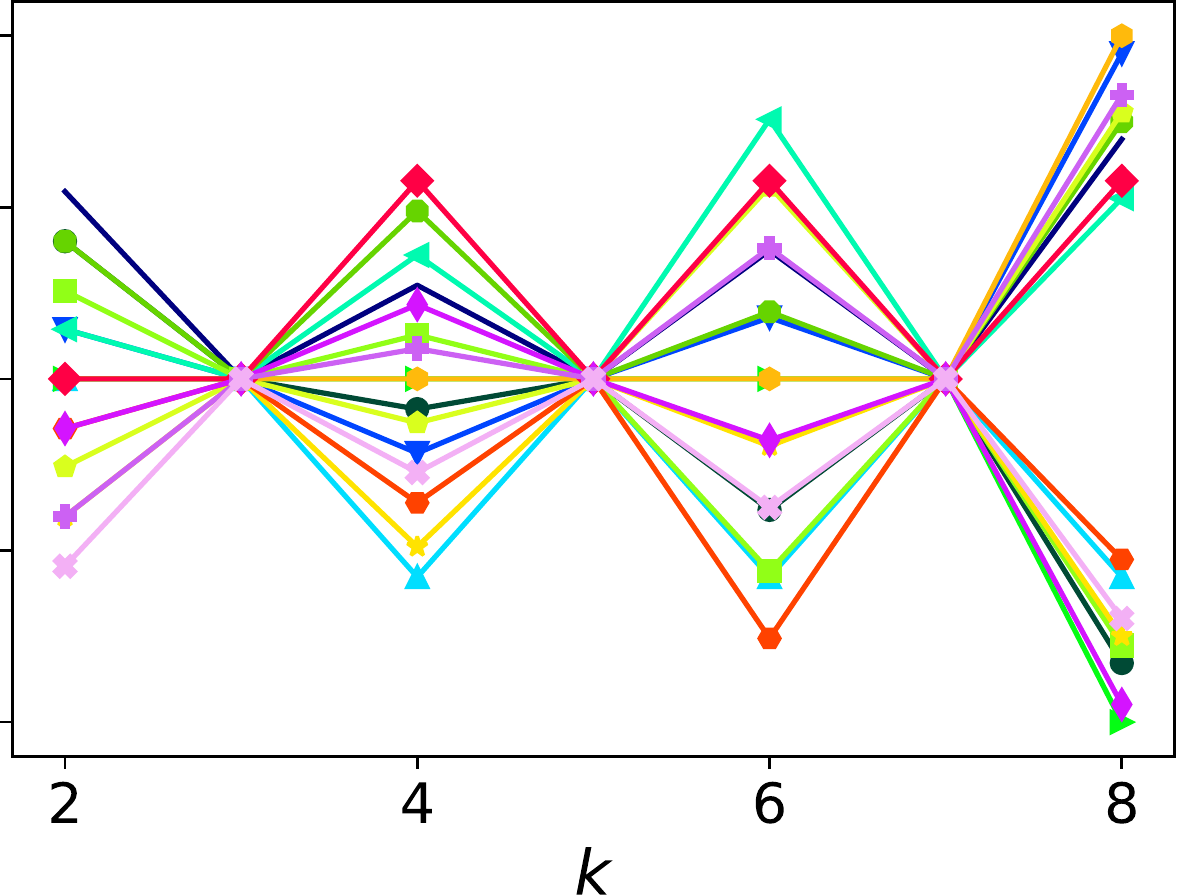}}
		\end{minipage}\begin{minipage}{0.33\textwidth}	
			\subfloat{(c) $\mu=2$}{	
				
				\includegraphics[height=3.7cm]{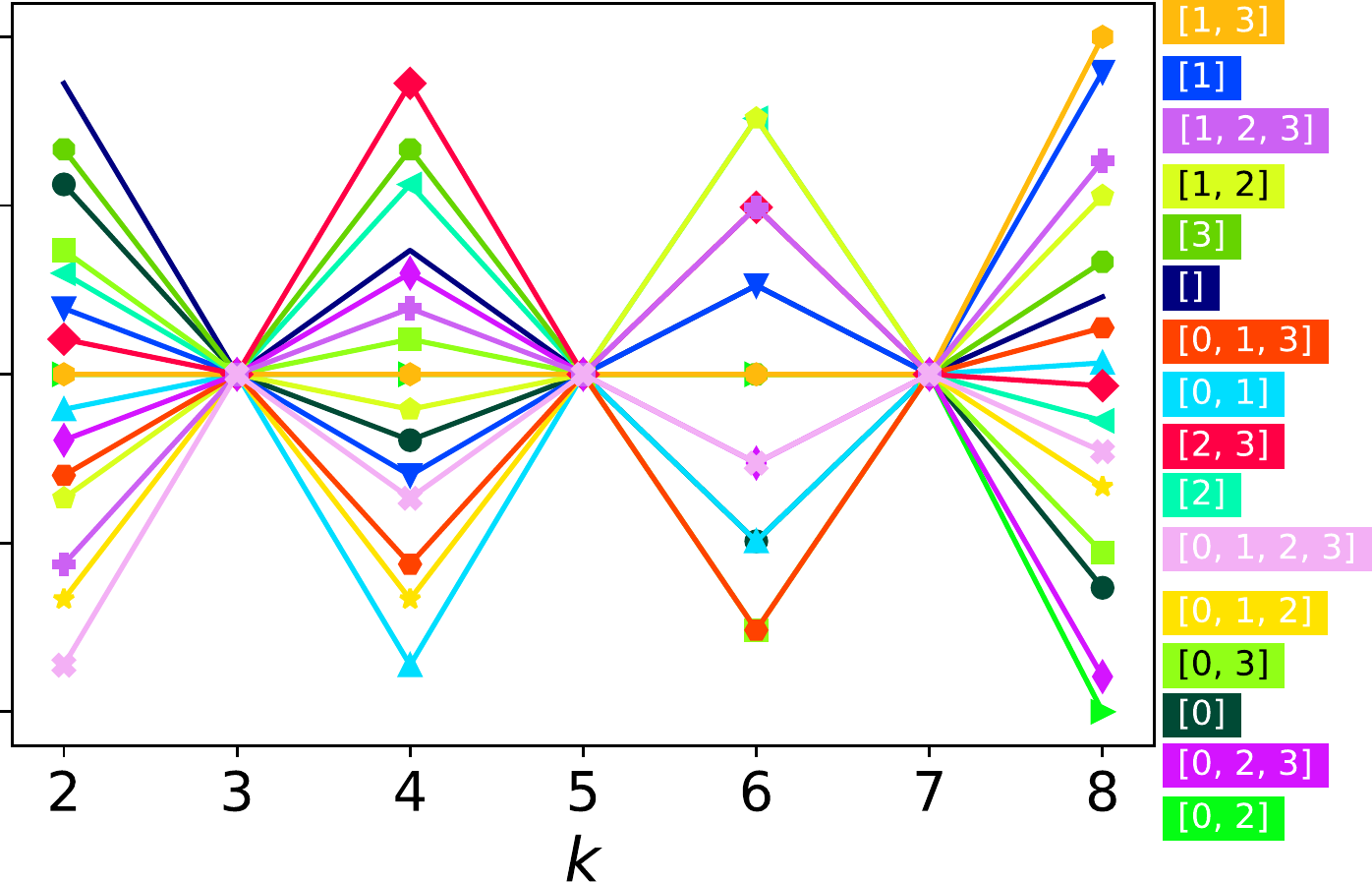}}
			
		\end{minipage}
		\caption{Site correlation function where $k$ denotes the site at different values of the chemical potential in the topological regime $\Delta=-t=1$ obtained by fermionic Gaussian states. The colour code is explained on the right.\label{fig:corr4}}
	\end{figure*}

\end{document}